\documentstyle[11pt,aaspp4,epsf,rotate]{article}
\def  \La          {\ifmmode {\rm Ly}\alpha \else Ly$\alpha$\fi}
\def  \Lalpha      {\ifmmode {\rm Ly}\alpha\,\lambda1215
		    \else Ly$\alpha$\,$\lambda1215$\fi}
\def  \Ka          {\ifmmode {\rm K}\alpha \else K$\alpha$\fi}
\def  \Lb          {\ifmmode {\rm L}\beta \else L$\beta$\fi}
\def  \Ha          {\ifmmode {\rm H}\alpha \else H$\alpha$\fi}
\def  \Halpha      {\ifmmode {\rm H}\alpha\,\lambda6563
		    \else H$\alpha$\,$\lambda6563$\fi}
\def  \Hb          {\ifmmode {\rm H}\beta \else H$\beta$\fi}
\def  \Hbeta       {\ifmmode {\rm H}\beta\,\lambda4861
		    \else H$\beta$\,$\lambda4861$\fi}
\def  \Pa          {\ifmmode {\rm P}\alpha \else P$\alpha$\fi}
\def  \HeI         {\ifmmode {\rm He}\,{\sc i}\,\lambda5876
		    \else He\,{\sc i}\,$\lambda5876$\fi}
\def  \HeII        {\ifmmode {\rm He}\,{\sc ii}\,\lambda1640
		    \else He\,{\sc ii}\,$\lambda1640$\fi}
\def  \CIII        {\ifmmode {\rm C}\,{\sc iii}\,\lambda977
		    \else C\,{\sc iii}\,$\lambda977$\fi}
\def  \CIIIb       {\ifmmode {\rm C}\,{\sc iii]}\,\lambda1909
		    \else C\,{\sc iii]}\,$\lambda1909$\fi}
\def  \CIV         {\ifmmode {\rm C}\,{\sc iv}\,\lambda1549
		    \else C\,{\sc iv}\,$\lambda1549$\fi}
\def  \bOIIIbA     {\ifmmode {\rm [O}\,{\sc iii]}\,\lambda4363
		    \else [O\,{\sc iii]}\,$\lambda4363$\fi}
\def  \bOIIIbB     {\ifmmode {\rm [O}\,{\sc iii]}\,\lambda5007
		    \else [O\,{\sc iii]}\,$\lambda5007$\fi}
\def  \OIIIb       {\ifmmode {\rm O}\,{\sc iii]}\,\lambda1663
		    \else O\,{\sc iii]}\,$\lambda1663$\fi}
\def  \OVb         {\ifmmode {\rm O}\,{\sc v]}\,\lambda1218
		    \else O\,{\sc v]}\,$\lambda1218$\fi}
\def  \OVI         {\ifmmode {\rm O}\,{\sc vi}\,\lambda1035
		    \else O\,{\sc vi}\,$\lambda1035$\fi}
\def  \OIVb        {\ifmmode {\rm O}\,{\sc iv]}\,\lambda1402
		    \else O\,{\sc iv]}\,$\lambda1402$\fi}
\def  \bOIIb       {\ifmmode {\rm [O}\,{\sc ii]}\,\lambda3727
		    \else [O\,{\sc ii]}\,$\lambda3727$\fi}
\def  \bOIb        {\ifmmode {\rm [O}\,{\sc i]}\,\lambda6300
		    \else [O\,{\sc i]}\,$\lambda6300$\fi}
\def  \NV          {\ifmmode {\rm N}\,{\sc v}\,\lambda1240
		    \else N\,{\sc v}\,$\lambda1240$\fi}
\def  \NIVb        {\ifmmode {\rm N}\,{\sc iv]}\,\lambda1486
		    \else N\,{\sc iv]}\,$\lambda1486$\fi}
\def  \NIIIb       {\ifmmode {\rm N}\,{\sc iii]}\,\lambda1750
		    \else N\,{\sc iii]}\,$\lambda1750$\fi}
\def  \MgII        {\ifmmode {\rm Mg}\,{\sc ii}\,\lambda2798
		    \else Mg\,{\sc ii}\,$\lambda2798$\fi}
\def  \bNeVb       {\ifmmode {\rm [Ne}\,{\sc v]}\,\lambda3426
		    \else [Ne\,{\sc v]}\,$\lambda3426$\fi}
\def  \NeVIII      {\ifmmode {\rm Ne}\,{\sc viii}\,\lambda774
		    \else Ne\,{\sc viii}\,$\lambda774$\fi}
\def  \SiIV        {\ifmmode {\rm Si}\,{\sc iv}\,\lambda1397
		    \else Si\,{\sc iv}\,$\lambda1397$\fi}
\def  \bFeXb       {\ifmmode {\rm [Fe}\,{\sc x]}\,\lambda6734
		    \else [Fe\,{\sc x]}\,$\lambda6734$\fi}
\def  \bFeXIb      {\ifmmode {\rm [Fe}\,{\sc xi]}\,\lambda7892
		    \else [Fe\,{\sc xi]}\,$\lambda7892$\fi}
\def  \FeII        {\ifmmode {\rm Fe}\,{\sc ii}\,
		    \else Fe\,{\sc ii}\,\fi}

\def \bSiIXbA  {\ifmmode {\rm [Si}\,{\sc ix]}\,\lambda 2.6 \mu{\rm m}
		    \else [Si\,{\sc ix]}\,$\lambda 2.6 \mu$m\fi}
\def \bSiIXbB  {\ifmmode {\rm [Si}\,{\sc ix]}\,\lambda 3.9 \mu{\rm m}
		    \else [Si\,{\sc ix]}\,$\lambda 3.9 \mu$m\fi}
\def \bMgVIIIb {\ifmmode {\rm [Mg}\,{\sc viii]}\,\lambda 3.0 \mu{\rm m}
		    \else [Mg\,{\sc viii]}\,$\lambda 3.0\mu$m\fi}
\def \bMgIVb   {\ifmmode {\rm [Mg}\,{\sc iv]}\,\lambda 4.5 \mu{\rm m}
		    \else [Si\,{\sc iv]}\,$\lambda 4.5 \mu\mu$m\fi}
\def \bNeVIb   {\ifmmode {\rm [Ne}\,{\sc vi]}\,\lambda 7.6 \mu{\rm m}
		    \else [Ne\,{\sc vi]}\,$\lambda 7.6\mu$m\fi}
\def \bArIIIb  {\ifmmode {\rm [Ar}\,{\sc iii]}\,\lambda 9.0 \mu{\rm m}
		    \else [Ar\,{\sc iii]}\,$\lambda 9.0\mu$m\fi}
\def \bSIVb    {\ifmmode {\rm [S}\,{\sc iv]}\,\lambda 10.5 \mu{\rm m}
		    \else [S\,{\sc iv]}\,$\lambda 10.5\mu$m\fi}
\def \bNeIIb   {\ifmmode {\rm [Ne}\,{\sc ii]}\,\lambda 12.8 \mu{\rm m}
		    \else [Ne\,{\sc ii]}\,$\lambda 12.8\mu$m\fi}
\def \bNeVbA   {\ifmmode {\rm [Ne}\,{\sc vi]}\,\lambda 14.3 \mu{\rm m}
		    \else [Ne\,{\sc vi]}\,$\lambda 14.3\mu$m\fi}
\def \bNeIIIbA {\ifmmode {\rm [Ne}\,{\sc iii]}\,\lambda 15.6 \mu{\rm m}
		    \else [Ne\,{\sc iii]}\,$\lambda 15.6\mu$m\fi}
\def \bSIIIbA  {\ifmmode {\rm [S}\,{\sc iii]}\,\lambda 18.7 \mu{\rm m}
		    \else [S\,{\sc iii]}\,$\lambda 18.7\mu$m\fi}
\def \bNeVbB   {\ifmmode {\rm [Ne}\,{\sc vi]}\,\lambda 24.3 \mu{\rm m}
		    \else [Ne\,{\sc vi]}\,$\lambda 24.3\mu$m\fi}
\def \bOIVb    {\ifmmode {\rm [O}\,{\sc iv]}\,\lambda 25.9 \mu{\rm m}
		    \else [O\,{\sc iv]}\,$\lambda 25.9\mu$m\fi}
\def \bSIIIbB  {\ifmmode {\rm [S}\,{\sc iii]}\,\lambda 33.5 \mu{\rm m}
		    \else [S\,{\sc iii]}\,$\lambda 33.5\mu$m\fi}
\def \bSiIIb   {\ifmmode {\rm [Si}\,{\sc ii]}\,\lambda 34.8 \mu{\rm m}
		    \else [Si\,{\sc ii]}\,$\lambda 34.8\mu$m\fi}
\def \bNeIIIbB {\ifmmode {\rm [Ne}\,{\sc iii]}\,\lambda 36.0 \mu{\rm m}
		    \else [Ne\,{\sc iii]}\,$\lambda 36.0\mu$m\fi}
\def \EBV {\ifmmode E_{\sc b-v} \else $E_{\sc b-v}$ \fi}
\def \Fnlr {\ifmmode F_{\sc nlr} \else $F_{\sc nlr}$ \fi}
\def \lgs {\ifmmode \log^2S \else $\log^2S$ \fi}
\def \qion {\ifmmode q_{\rm ion} \else $q_{\rm ion}$ \fi}
\def \tion {\ifmmode \theta_{\rm ion} \else $\theta_{\rm ion}$ \fi}
\def \ml {\ifmmode m_\ell \else $m_\ell$ \fi}
\def \fl {\ifmmode f_\ell \else $f_\ell$ \fi}
\def \kl {\ifmmode k_\ell \else $k_\ell$ \fi}
\def \Eion {\ifmmode E_{\rm ion} \else $E_{\rm ion}$ \fi}
\def \arcsec {^{\prime\prime}}
\def \maxs {\ifmmode \max S_\ell \else $\max S_\ell$ \fi}
\def \Mo {\ifmmode M_\odot \else $M_\odot$ \fi}
\def \Ro {\ifmmode R_\odot \else $R_\odot$ \fi}
\def \Lo {\ifmmode L_\odot \else $L_\odot$ \fi}
\lefthead{Alexander et al.}
\righthead{The ionizing continuum of AGN NGC\,4151}

\begin{document}

\title{Infrared spectroscopy of NGC\,4151:\\
       Probing the obscured ionizing AGN continuum}

\author{
Tal Alexander\altaffilmark{1,2,3}, 
Eckhard Sturm\altaffilmark{1},
Dieter Lutz\altaffilmark{1},
Amiel Sternberg\altaffilmark{2},
Hagai Netzer\altaffilmark{2},
and
Reinhard Genzel\altaffilmark{1}
}

\altaffiltext{1}{Max-Planck-Institut f\"ur Extraterrestrische Physik, 
Postfach 1603, D-85740 Garching, Germany}
\altaffiltext{2}{School of Physics and Astronomy and Wise Observatory, 
Raymond and Beverly Sackler Faculty of Exact Sciences,
Tel Aviv University, Ramat Aviv, Tel Aviv 69978, Israel}
\altaffiltext{3}{Institute for Advanced Study, Olden Lane, 
Princeton, NJ 08540, USA}

\begin{abstract}
The ISO-SWS\footnote{Based on observations made with ISO, an ESA
project with instruments funded by ESA member states (especially the
PI countries: France, Germany, The Netherlands and the United Kingdom)
and with the participation of ISAS and NASA. The SWS is a joint
project of SRON and MPE.} infrared spectroscopic observations of the
nucleus of Seyfert galaxy NGC\,4151, which are described in a
companion paper, are used together with a compilation of UV to IR
narrow emission line data to determine the spectral shape of the
obscured extreme-UV continuum that photoionizes the narrow line
emitting gas in the active galactic nucleus. We present a new method
to determine the best fitting photoionizing continuum and emission
line cloud model from a heterogeneous set of emission line data.

For NGC\,4151, we find a best-fit model that reproduces the observed
line fluxes to within a factor of 2 on average, and which is
consistent with the observed geometry of the optical Narrow Line
Region (NLR). Our model consists of a clumpy, optically thick
(ionization bounded) gas distribution, with a hydrogen gas density of
$\sim 1000$\,cm$^{-3}$ and a volume filling factor of $6.5\,10^{-4}$.

Our best fitting spectral energy distribution (SED) falls sharply
beyond the Lyman limit and then rises sharply again towards 100 eV. In
particular, it does not display a `Big Blue Bump' signature of a hot
accretion disk. We find that this SED, which best reproduces the NLR
line emission on the 100--500 pc scale, does not have enough UV
photons to produce the observed BLR recombination emission from the inner 1
pc. This suggests that the BLR is photoionized by the intrinsic
continuum source, which does have a strong UV component (perhaps a Big
Blue Bump), but that this UV component is absorbed by material located
between the NLR and BLR. Our analysis suggests that the absorber
consists of $\sim 5\,10^{19}$\,cm$^{-2}$ of neutral hydrogen. Such an
absorber was independently detected by UV absorption lines (Kriss et
al. \cite{Kriss,KDZKE}).

Using our new method, we confirm our previous conclusion that a Big
Blue Bump is present in the SED of the Seyfert 2 galaxy Circinus.
\end{abstract}

\keywords{galaxies: individual NGC\,4151 --- galaxies: Seyfert ---
          quasars: emission lines --- infrared: ISM: lines and bands}

\section{Introduction}

The intrinsic spectral energy distribution (SED) of active galactic nuclei
(AGNs), which extends from the radio up to $\gamma$-rays, cannot be observed
from the Lyman limit and up to several hundred eV due to Galactic and
intrinsic absorption. It is widely assumed that the intrinsic SED in this
range is a superposition of a bump and one or more power-law components
(Figure~\ref{f:sed+bump}). The hypothetical bump is known as the `Big Blue
Bump'. The deviation from the hypothetical power-law at the high energy
wing of the bump is known as the `soft excess'. This latter feature
receives special attention because it may be within the observable soft
X-ray range.

There are theoretical and observational motivations for the bump plus
power-law hypothesis. Generically, thermal radiation from accretion on a
black hole has a typical effective temperature $T_a$ that is related to the
typical size of the emitting region $r_a$ and to the accretion luminosity
$L_a$ by $L_a \propto 4\pi r_a^2 T_a^4$.  Order of magnitude estimates, as
well as more detailed accretion disk models, suggest that typical AGN
accretion temperatures should be of order ${\rm few}\times10^5$\,K,
corresponding to a Wien bump peaking at $\sim 40$\,eV (see e.g. Blandford
\cite{BNW} for a review). However, thermal bumps may be absent if the
accretion is advection-dominated (Narayan \& Yi
\cite{NY}; Narayan, Kato \& Fumio \cite{NKH}).

The power-law components are suggested by the observed shape of the SED at
the optical-UV and soft to hard X-ray continua, which in many cases have
different spectral indices. The power-law continua are thought to be
produced by inverse Compton scattering of accretion disk
emission. Observationally, there are indications that the mean optical--UV
slope tends to be flatter than the mean soft X-ray slope (e.g. Puchnarewicz
et al. \cite{Puchnarewicz}). This suggests a break somewhere in between
these two spectral ranges, on top of which a bump may be superimposed.

Two major issues motivate the efforts to uncover the ionizing SEDs of
AGNs. The AGN paradigm postulates that AGNs are powered by an accreting
super-massive black hole. Major advances have been made in recent years in
establishing the existence of super-massive black holes in the nuclei of
normal and active galaxies (Kormendy \& Richstone \cite{KR}; Miyoshi et al
\cite{Miyoshi}; Eckart \& Genzel \cite{EG}). On the other hand, there is
little direct evidence to support the accretion hypothesis. A detection of
a signature of the accretion mechanism in the SED will have important
implications both for validating the AGN paradigm as well as for
discriminating between various accretion models. The ionizing SED is also
important for modeling the effects of the AGN on its environment, since the
SED is a critical input for photoionization models. Such modeling is a
major tool in the study of AGN emission lines and in attempts to
disentangle the contribution of the nuclear emission from that of
starbursting regions in ultra-luminous IR galaxies (Lutz et
al. \cite{Lutz1}; Genzel et al. \cite{Genzel}).

The spectral range of the SED that is accessible to direct observations can
be maximized in the soft X-ray regime by observing nearby Seyfert galaxies,
and in the UV by observing high redshift AGNs (e.g. Zheng et
al. \cite{Zheng}), although this is limited by inter-galactic
absorption. Even so, a large gap remains. Almost all the current
information on the SED in this gap comes from combining UV and X-ray data,
following two basic approaches. The first is to interpolate, assuming some
model, between the UV and X-rays. The second is to look for correlations
between the continuum slopes at the optical-UV and X-rays.

A survey of results from recent X-ray studies (Walter \& Fink \cite{WF};
Laor et al. \cite{LFEWM1}; Walter et al. \cite{WOCFMOW}; Puchnarewicz et
al. \cite{Puchnarewicz}; Laor et al. \cite{LFEWM2}; Brunner et
al. \cite{Brunner}) reveals a confusing picture. Answers to the key
observational questions: Is there a bump? Is there a soft excess? What is
the object-to-object scatter in bump temperature? Is intrinsic absorption
significant?, vary greatly. As a consequence, answers to key theoretical
questions, such as the consistency of the SEDs with accretion models and
the origin of the object-to-object scatter, also remain inconclusive. There
are several possible reasons for this. One is the strong and rapid
variability of AGNs in the X-rays. To date, very few AGNs have been observed
simultaneously in the the UV and X-rays (NGC\,4151 is one of them). Another
problem is the very low spectral resolution of current X-ray spectroscopy,
which requires a-priori models of the SED and of the galactic absorption to
reconstruct the spectrum. As a result, the conclusions are strongly model
dependent and many different models fit the data equally well (see
e.g. Walter et al. \cite{WOCFMOW}). Because of this ambiguity, the
interpolation approach has been primarily applied to samples of AGNs rather
than to individual objects. The correlation approach is, by definition,
statistical and applicable only to AGN samples. Therefore, both approaches
are vulnerable to the statistical pitfalls of sample analysis, such as
selection effects and small sample fluctuations. It is widely suspected
that selection effects play an important role in biasing the results of
X-ray sample studies (e.g. Puchnarewicz et al. \cite{Puchnarewicz}). If
this is true, then there must be a large variety of AGN SEDs. This raises
doubts whether the sample mean is a useful quantity for describing such a
heterogeneous population. In this situation, a method that can be applied
to an {\em individual object} rather than to AGN {\em samples} can be
extremely useful. It is in this that IR spectroscopy can play a role.

Following the first detection of $\bFeXb$ in NGC\,4151 (Oke \& Sargent
\cite{OS}), it became apparent that AGNs emit optical lines from highly
ionized species (`coronal lines').  More recently, IR coronal lines have
been detected with ground based telescopes (Oliva \& Moorwood \cite{OM},
Oliva et al. \cite{Oliva}, Reconditi \& Oliva \cite{RO}), and
photoionization models predicted that many additional lines were likely to
be observed in the spectral range covered by ISO (e.g. Spinoglio \& Malkan
\cite{SM}, Greenhouse et al. \cite{Greenhouse}, Voit \cite{Voit}). The
observed line ratios point to photoionization rather than collisional
ionization as the dominant ionization mechanism (Oliva et
al. \cite{Oliva}). In this case, each of these lines probes the SED at
energies $\ge \Eion$, where we designate by $\Eion$ the ionization
energy required to produce the emitting ion from the preceding
ionization stage. Figure~\ref{f:eion} demonstrates that the gap in
the SED of NGC\,4151 is well covered by the observed optical and IR
lines.  Therefore, by fitting a photoionization model to the observed
lines, it should be possible to constrain the SED.

With this in mind, our ISO-SWS program on bright galactic nuclei
included detailed spectroscopic observations of several AGNs. The
results from a pilot study of the nearby Seyfert 2 Circinus Galaxy
(Moorwood et al. \cite{Moorwood}) demonstrated the potential of this
approach to the reconstruction of the ionizing SED. Here we extend
this method to NGC\,4151, one of the brightest, closest and most
extensively studied Seyfert galaxies. The ISO-SWS observations of
NGC\,4151 are presented in a companion paper (Sturm et
al. \cite{Sturm}, Paper I), which compares the IR emission line
profiles to optical line profiles and derives constraints on models of
profile asymmetry. In this work, we add observed UV, optical, near-IR
and far-IR lines from the literature to the mid-IR SWS lines, and
perform an extensive search in parameter space to find the best
fitting SED.

The rest of this paper is organized as follows. The observed properties of
Seyfert galaxy NGC\,4151 and the inferences about its physical conditions
are discussed in Sect.~\ref{s:n4151}. The observed emission line fluxes
compilation and the criteria for selecting the lines to use in the modeling
are described in Sect.~\ref{s:lines}. The photoionization models and the
way they are compared to the data are described in
Sect.~\ref{s:models}. The results are presented in Sect.~\ref{s:results},
discussed in Sect.~\ref{s:discuss} and summarized in Sect.~\ref{s:summary}.

\section{The physical properties of NGC\,4151}
\label{s:n4151}

Although NGC\,4151 is often called the `classical' Seyfert galaxy, it
displays a variety of properties typical of different AGN classes. It
may therefore be in fact more complex than average. We present here a
brief overview of its properties that are relevant to this work.

NGC\,4151 is a barred spiral galaxy, which is seen almost face on
(Pedlar et al. \cite{Pedlar92}) at redshift $z=0.0033$ (distance
$D=9.9h^{-1}$\,Mpc) and magnitude $m_V = 11.5$ (Brinkmann et
al. \cite{Brinkmann}). Its spectrum is dominated by the non-stellar
component (Kaspi et al. \cite{Kaspi}). NGC\,4151 is highly variable in
both the continuum and the broad emission lines. Originally classified
as an intermediate, Seyfert 1.5 galaxy (Osterbrock \& Koski
\cite{OK}), it went through an extreme low state in 1984, where it
took on the characteristics of a Seyfert 2 galaxy (Penston \& P\'erez
\cite{PP}). Currently, NGC\,4151 is usually grouped in the Seyfert 1 class.

\subsection{The ionizing continuum}
\label{s:n4151-SED}

The December 1993 multi-wavelength campaign to monitor this variability
(Crenshaw et al. \cite{Crenshaw}; Kaspi et al. \cite{Kaspi}; Warwick et
al. \cite{Warwick}; Edelson et al. \cite{Edelson}) yielded a simultaneous
SED extending from the optical to $\gamma$-rays. NGC\,4151 was observed
in the optical from Wise Observatory and Lowell Observatory, in the UV by
IUE, in X-rays by ROSAT and ASCA and in $\gamma$-rays by
CGRO. Warwick et al. (\cite{Warwick}) deconvolved the X-rays to $\gamma$-rays
spectrum by assuming an $F_\nu\propto\nu^{-0.5}$ powerlaw and a thermal
bremsstrahlung component with a temperature of 0.5 keV, absorbed by a two
component intrinsic absorber (`partial covering model') and by galactic
neutral hydrogen.

In order to extend the spectral coverage up to the Lyman limit, we
used the 1995 HUT best-fit power-law continuum (Kriss et
al. \cite{KDZKE}), which we added to the 1993 SED after normalizing
them to the IUE flux at the line-free wavelength of 1806\AA\,. A large
uncertainty is associated with this flux value due to variability and
reddening. The 1995 UV continuum was $\sim5$ times more luminous and
much bluer than it was in 1990 (Kriss et al. \cite{Kriss}). By fitting
a power-law continuum to the HUT data, Kriss et al. (\cite{Kriss})
derive $\EBV=0.04$, which they use to deredden the SED. This amounts
to a factor of 1.3 increase in the flux at 1806\AA\,. However,
estimates of the reddening vary greatly (Sect. \ref{s:n4151-gas}). In
view of these problems, we treat the flux at 1 Ryd as another source
of uncertainty in the model, which is investigated by the model
fitting procedure described in Sect.~\ref{s:fit}.

Figure \ref{f:sed+bump} shows the composite SED, where for compatibility
with Edelson et al. (\cite{Edelson}), we plot the SED as $L_\nu = 4\pi D^2
f_\nu$, where $f_\nu$ is the observed flux density in
erg\,s$^{-1}$\,cm$^{-2}$\,Hz$^{-1}$ and $D=20$\,Mpc. The SED displays the
typical strong optical-UV emission lines of a Seyfert galaxy and
significant intrinsic absorption below 4 keV. Above 4 keV, a hard power-law
extends up to a cutoff at $\sim90$\,keV. Also visible are X-rays emission
lines at $\sim 6.4$\,keV. Figure \ref{f:sed+bump} also shows a schematic
sketch of a superimposed Big Blue Bump and a powerlaw spectrum, constructed
by superimposing a power law spectrum, $L_\nu\propto\nu^{-1}$ and 5
blackbody spectra normalized to have equal total luminosities,
$B_\nu(T)/T^4$, with $T=10^5, 2\,10^5,\ldots,5\,10^5$\,K. Note that the
shape of the excess emission above the power-law depends in part on the way
the SED is presented. A bump in $\nu L_\nu$ may appear only as a `shoulder'
or a flattening in $L_\nu$. Here, we use the term `bump' irrespective of
the presentation of the SED.

\subsection{The galactic nucleus}
\label{s:n4151-nucleus}
The line spectrum of NGC\,4151 points to the existence of the three
kinematically distinct nuclear emission regions that are observed in many
Seyfert galaxies: an unresolved Broad Line Region (BLR) with typical line
widths of $> 5000$\,km s$^{-1}$, densities above $10^8$\,cm$^{-3}$ and an
ionization parameter (see below) $U \sim 0.03$--$1$, a Narrow Line Region
(NLR) with typical line widths of $\la 500$\,km s$^{-1}$, densities in the
range $10^3$ to $10^6$\,cm$^{-3}$ and $U \sim 0.001$--$0.01$, and an
Extended Emission Line Region (EELR) with typical line widths of $\la 50$
km s$^{-1}$, densities below $10^3$\,cm$^{-3}$ and $U \la 0.005$ (Schulz \&
Komossa \cite{SK}). The line widths of the SWS forbidden IR lines point to
an origin in the NLR. This is supported by estimates of the gas density
from the line ratios (Sect. \ref{s:n4151-gas}) and is consistent with the
small critical densities ($<{\rm few}\times10^5$\,cm$^{-3}$) of the mid-IR
fine-structure line transitions, which are incompatible with the
BLR. However, the very large aperture of the SWS includes also the EELR,
and some contribution from the EELR cannot be ruled out, especially for the
lower ionization lines.

Narrow band images in the [O\,{\sc ii}], [O\,{\sc iii}] and Balmer lines
(Heckman \& Balick \cite{HB}; P\'erez et al. \cite{Perez}; Unger et
al. \cite{Unger}; P\'erez-Fournon \& Wilson \cite{PW}; Yoshida \& Ohtani
\cite{YO}; Hutchings et al. \cite{Hutchings}) show that NGC\,4151 has an
elongated, knotty EELR which extends along PA $\sim 50^\circ$ up to $\sim
30\arcsec$ to the SW of the nucleus, but only up to $\sim 15\arcsec$ to the
NE. The NLR symmetry axis coincides with the radio jet at PA
$\sim77^\circ$ (Pedlar el al. \cite{Pedlar93}), but is misaligned with the
EELR.  This misalignment is consistent with a very wide ionization cone
with an opening angle $\sim 120^\circ$, intercepting the galactic disk at
grazing incidence (Pedlar et al. \cite{Pedlar92},
\cite{Pedlar93}). Images show that the opening angle of the line
emitting conical section of the galactic plane is $\sim 80^\circ$
(Hutchings et al. \cite{Hutchings}). The $\bOIIIbB$ profile, which traces
the velocity field (Schulz \cite{Schulz90}) and the $\bOIIIbB/\Ha$ line
ratio, which traces $U$ (Robinson et al. \cite{Robinson}), both indicate a
NLR size of $\sim5\arcsec$. An $\bOIIIbB$ map of the inner nucleus reveals
NLR emission from $\la 0.5\arcsec$ of the nucleus up to $2-3\arcsec$
(Hutchings et al. \cite{Hutchings}).  These size estimates are unlikely to
be affected significantly by projection, since the inclination angle of the
galactic disk at the inner few arcsec is estimated at
$12^\circ<i<21^\circ$, and the NLR symmetry axis is within a few degrees of
the line of nodes (Boksenberg et al. \cite{Boksenberg}).
  
The volume filling factor of the NLR gas, $F$, which is
defined as the fraction of the volume actually filled with gas\footnote{By
definition, $F^{1/2} = n_{\rm rms}/n_{\rm loc}$, the ratio of the volume
averaged rms electron density to that of the local one. $n_{\rm loc}$ is
estimated from emission line ratios, while $n_{\rm rms}$ is estimated from
the observed flux of a recombination line, $\fl$, and the angular size of
the line emitting region,$\tion$, from $4\pi D^2 \fl \propto n_{\rm rms}^2
(D\tion)^3$.  This introduces an $h$ dependence to the estimate of the
filling factor $F=F_1h$.}, is very small. Yoshida \& Ohtani (\cite{YO})
estimate $F \sim 10^{-3}h$ based on the low excitation $\Ha$, $\bOIIb$ and
$\bOIIIbB$ maps. Similarly, Robinson et al. (\cite{Robinson}) find $F \sim
{\rm few}\times 10^{-4}h$ in an $\Ha$ knot at the edge of the NLR. Thus the
low excitation NLR lines are emitted from a very clumpy distribution of
gas. Unfortunately, narrow band images or position-resolved long-slit
spectroscopy of coronal lines are not yet available for NGC\,4151, so their
NLR size and filling factor cannot be estimated directly. However, an upper
limit on the size of the coronal line emitting region can be estimated from
the low ionization NLR lines with the plausible assumption that the high
ionization emission is more compact. This is indeed observed in the spatial
distribution of the $\CIV$ line relative the $\bOIIIbB$ line (Hutchings et
al. \cite{Hutchings}), and is observed in the coronal lines of the Circinus
galaxy (Maiolino et al. \cite{Maiolino}).

Spatially resolved maps of the line emission in the NLR and EELR make it
possible to estimate the anisotropy parameter $A$, defined as the ratio
between the ionizing luminosity that the gas must see in order to emit the
observed line flux, and that directed at the observer. There are
indications that the anisotropic bi-conic structure of the NLR and EELR
traces an anisotropy in the photoionizing luminosity (as opposed to an
anisotropy in the matter distribution). Estimates of $A$ vary greatly since
they strongly depend on the assumed intrinsic continuum (usually
interpolated over the gap in the SED by a power-law) and the
photoionization model of the NLR /EELR. Schulz \& Komossa (\cite{SK})
derive $1\le A \le 6$, Yoshida \& Ohtani (\cite{YO})) derive $1<A\la 3$,
Penston et al. (\cite{Penston90}) derive $A\sim13$, and Robinson et
al. (\cite{Robinson}) obtain $A\sim1$ and $A\sim10$ by two different
methods.

Strong absorption on the line of sight in the optical, UV and X-ray
bands implies the existence of a multi-component system of absorbers
that may also modify the ionizing SED in other directions. Warwick et
al. (\cite{Warwick}) fit the X-ray SED with a two component absorber
of total hydrogen column density in the range $10^{22}$ to $10^{23}$
cm$^{-2}$. The location of this absorber along the line of sight is
unknown.  Kriss et al. (\cite{Kriss}) fit the Lyman absorption lines
with an outflowing clumpy and dense absorber ($n > 10^{9.5}$
cm$^{-3}$) with a neutral hydrogen column density in the range
$6\,10^{17}$ to $6\,10^{20}$\,cm$^{-2}$, which covers a large fraction
of the continuum source and the BLR along the line of sight. Both the
high density and the outflow velocity suggest that the absorber is
between the BLR and NLR. The properties of this absorber are
incompatible with that of the X-ray absorber, and it must be assumed
that these are two different components. HUT observations of NGC\,4151
at a later epoch (Kriss et al. \cite{KDZKE}) show that the properties
of the absorber appear to change in time, possibly in response to
changes in the continuum luminosity, and indicate the possible
presence of an additional component with neutral hydrogen column
density of up to $5\,10^{20}$\,cm$^{-2}$.

\subsection{The line emitting gas}
\label{s:n4151-gas}

   The ISO spectra can be used to estimate the gas density and
temperature, given estimates of the reddening and starburst
contribution to the line emission.

Estimates of the reddening of the NLR in NGC\,4151 range from almost
negligible, $\EBV = 0.04$ (Kriss et al. \cite{KDZKE}), 0.05 (Penston et
al. \cite{Penston81}; Wu \& Weedman \cite{WW}; Boksenberg et
al. \cite{Boksenberg78}), to considerable, $\EBV = 0.12$ (Kaler
\cite{Kaler}), 0.13 (Malkan \cite{Malkan}), and 0.09 to 0.28 (Ward et
al. \cite{Ward}). The elemental abundances and the internal dust content of
the line emitting gas are even less certain. In a study of the Circinus
galaxy, Oliva et al (\cite{Oliva}) find that the coronal line emitting gas
is dust free and has solar abundances to within a factor of $\sim 2$.
Similarly, Ferguson et al. (\cite{FKF}) compare an extensive grid of
coronal line photoionization models to observed line ratios in several
Seyfert galaxies and conclude that dust is unlikely to exist in coronal
line emitting gas, and that the Si and Ca abundances are solar to within a
factor of $\sim2$. The NGC\,4151 EELR models of Schulz \& Komossa
(\cite{SK}) point to a $\sim 1/2$ solar metallicity for all metals
(consistent with metal poor gas, but not with dust depletion). In our
models, we will assume that the abundances are solar and that the gas is
dust free.

The stellar population in the nucleus of NGC\,4151 can affect the
observed line fluxes in two ways. First, the presence of
star-formation regions within the large ISO-SWS aperture can introduce
line emission due to radiation fields peaking at $\sim1$\,Ryd, which
are typical of hot stars (if these regions are shielded from the
central continuum). An upper limit on the starburst contribution to
the lower ionization lines can be estimated from the upper limits on
the PAH emission, which are tracers of star forming activity (Roche et
al. \cite{RASW}; Genzel et al. \cite{Genzel}). The mean observed
ratios of the integrated $\bNeIIb$ line flux to that of the PAH
feature in starburst galaxies M83, NGC\,4945, NGC\,3256 and NGC\,7552
(see also Genzel et al. \cite{Genzel}) are $\bNeIIb$/PAH\,$\lambda
7.7\mu$m = 0.07 $\bNeIIb$/PAH\,$\lambda 6.2\mu$m = 0.22, as compared
to the NGC\,4151 lower limits of $\bNeIIb$/PAH\,$\lambda 7.7\mu{\rm m}
> 0.32$ and $\bNeIIb$/PAH\,$\lambda 6.2\mu{\rm m} > 0.47$, the latter
value being much less certain. The more reliable upper limit on the
PAH\,$\lambda 7.7\mu$m flux leads to an upper limit of $\sim 20$\% of
starburst contribution to $\bNeIIb$ in NGC\,4151.  This is supported
by the [O\,{\sc iii}], $\Ha$ and $\Hb$ line ratios and maps in the
bright knots of the EELR, which point to photoionization by the
central source rather than photoionization by hot stars (Schulz
\cite{Schulz88}; P\'erez et al. \cite{Perez}; P\'erez-Fournon \&
Wilson \cite{PW}). Second, a strong underlying stellar continuum may
affect the flux measurements of the optical lines. Robinson et
al. \cite{Robinson} fit the off-nuclear continuum ($5\arcsec$ SW of
the center) by an old stellar population and conclude that the effect
of absorption features on the emission line flux measurements is
negligible. Thus, stars are unlikely to affect the reconstruction of
the SED.

Three density-sensitive pairs of bright lines of the same ion are found
among the NGC\,4151 emission lines detected in the ISO-SWS wavelength
range: [Ne\,{\rm v}]\,$\lambda14.32,24.32\mu$m, [Ne\,{\sc
iii}]\,$\lambda15.55,36.04\mu$m, and [S\,{\sc
iii}]\,$\lambda18.71,33.48\mu$m. We also consider the [O\,{\sc
iii}]\,$\lambda51.81,88.35\mu$m pair observed with ISO-LWS by Spinoglio et
al. (\cite{Spinoglio}). Among these observations, the [Ne\,{\sc v}] ratio
is perhaps most interesting since it samples a NLR species with an
ionization energy near 100\,eV. However, for NGC\,4151 with its apparently
very small contribution of star forming activity to the emission line
spectrum, the [Ne\,{\sc iii}], [S\,{\sc iii}], and [O\,{\sc iii}] ratios
are valuable NLR diagnostics as well. Figure \ref{f:density} shows the
density dependence for these ratios, computed by solving the rate equations
for five level systems. Transition probabilities and collision strengths
have been taken from Mendoza \& Zeippen (\cite{MZ}) ([S\,{\sc iii}]),
Galavis et al. (\cite{GMZ95}) ([S\,{\sc iii}]), Galavis et
al. (\cite{GMZ97}) ([Ne\,{\sc iii}], [O\,{\sc iii}], [Ne\,{\sc v}]), Butler
\& Zeippen (\cite{BZ}) ([Ne\,{\sc iii}]), and Lennon and Burke (\cite{LB})
([Ne\,{\sc v}], [O\,{\sc iii}]).  ISO observations of the planetary nebula
NGC\,6302 (Pottasch et al. \cite{Pottasch}) support the Lennon \& Burke
collision strengths for [Ne\,{\sc v}] that were previously questioned by
Oliva et al. (\cite{OPR}).

The [O\,{\sc iii}] ratio of 1.5 measured by Spinoglio et
al. (\cite{Spinoglio}) for NGC\,4151 corresponds to a density of about
450\,cm$^{-3}$. The other ratios should be near their low density limits at
such densities and hence of little use to further constrain the density of
the NLR gas in NGC\,4151. Indeed, the measured ratios [Ne\,{\sc v}],
[Ne\,{\sc iii}], [S\,{\sc iii}] of 0.98, 5.9 and 0.67, respectively, are
close to their low density limits, and consistent with the [O\,{\sc iii}]
density. For simplicity and since [O\, {\sc iii}] may tend to sample
slightly lower density extended regions when compared to higher ionization
species, we adopt an electron density of 1000\,cm$^{-3}$ for our
modelling. This is still well below the critical densities of the lines we
model.
 
The infrared ground state fine-structure line fluxes are insensitive to the
electron gas temperature for $T\ga 2\,10^3$\,K. However, the optical
forbidden line intensities, which originate from higher energy levels, are
sensitive to the gas temperature of $\sim 10^4$\,K in photoionized gas. The
ratios of optical and infrared transition may therefore be useful
diagnostics of the gas temperature. The dominant uncertainty in such an
analysis is often the reddening correction to be applied to the optical
fluxes.  Because of the low extinction to the NLR of NGC\,4151, for which
we adopt $\EBV=0.05-0.15$ (see above), a temperature analysis based on such
an optical/infrared comparison appears promising. The four species
discussed above emit optical lines which are in principle suitable. The
optical/IR ratios of [S\,{\sc iii}]\,$\lambda 9096$\AA\, and [S\,{\sc
iii}]$\lambda 9531$\AA\, to $\bSIIIbA$ and $\bSIIIbB$ and of
$\bOIIIbA$\AA\, and $\bOIIIbB$\AA\, to [O\,{\sc iii}]\,$\lambda 51.81\mu$m
and [O\,{\sc iii}$\lambda 88.35\mu$m, however, are of limited diagnostic
value at the density of the NGC\,4151 NLR and electron temperatures in the
range 10000-20000\,K, since the temperature dependence is small compared to
the density dependence and compared to the effects of possible problems in
inter-calibration of optical and infrared measurements. In case of
[Ne\,{\sc v}], a suitable large aperture measurement of the 3426\AA\/
transition is not available. It is hence not possible to test claims of
very high $T_e$ in the highly ionized region that are based on analysis of
optical [Fe\,{\sc vii}] lines in several Seyferts (Erkens et
al. \cite{EAW}).

A robust diagnostic is the [Ne\,{\sc iii}]$\lambda3868$\AA\/ /
$15.55\mu$m ratio which is basically density-insensitive (Figure
\ref{f:temp}).  We adopt the 3868\AA\/ flux of Oke \& Sargent
(\cite{OS}) (Table \ref{t:flux}) and correct for
$\EBV=0.05-0.15$. Taking into acount this uncertainty in extinction,
and an additional 30\% calibration uncertainty which we have added
directly, we use the [Ne\,{\sc iii}] ratio to derive an electron
temperature of 13000$\pm$2500\,K, with lower electron temperature
obtained for lower adopted extinction. It is instructive to compare
this value to the classical NLR temperature diagnostic that uses the
optical [O\,{\sc iii}]$\lambda4363$\AA\/ / 5007\AA\/ ratio, since the
two species require similar ionization energies for their creation and
should sample a similar part of the NLR. Correcting the 4363\AA\/ /
5007\AA\/ ratio of 0.03 (Oke \& Sargent \cite{OS}) for
$\EBV=0.05-0.15$, we infer an [O\,{\sc iii}] electron temperature of
about 19000\,K. We caution however that this number may be very
uncertain due to the faintness of the 4363\AA\/ line. Similarly high
[O\,{\sc iii}] electron temperatures in NLRs and EELRs
(e.g. Storchi-Bergman et al. (\cite{SWMB}), using a more careful
treatment of the [O\,{\sc iii}]$\lambda4363$\AA\/ line) have been used
to put fairly strong constraints on photoionization scenarios
(e.g. Binette et al. \cite{BWS}), or to reject them entirely. An
electron temperature of 13000\,K as obtained from the optical/IR
[Ne\,{\sc iii}] ratio is however reproduced by fairly standard
photoionization models, thus avoiding the need for more complex
scenarios. Clearly, an extension of this analysis to other AGNs is
highly desirable to investigate this discrepancy.

\section{The observed line flux compilation}
\label{s:lines}

The advantage of the IR lines as diagnostics of the gas and SED lies in
their insensitivity to the gas temperature and to external
reddening. However, this also means that they cannot constrain these
physical parameters. In particular, UV and optical lines are needed to fix
$\EBV$.  In order to maximize the observational constraints on the models,
we added to the measured mid-IR lines observed UV, optical and NIR lines from
the literature, as well as recently measured far-IR lines.
We initially compiled a list of about 120 observed emission lines: UV
lines from Kriss et al. (\cite{Kriss}), optical lines from Oke \&
Sargent (\cite{OS}), Anderson (\cite{Anderson}), Netzer
(\cite{Netzer74}), Boksenberg et al. (\cite{BSAFPS}), Osterbrock \&
Koski (\cite{OK}); NIR lines from Osterbrock, Shaw \& Veilleux
(\cite{OSV}), Thompson (\cite{Thompson}). In addition, we included
some far-IR lines observed by the ISO Long Wavelength Spectrometer
(LWS) (Spinoglio et al. \cite{Spinoglio}).

Various problems stand in the way of merging these different
observations into a self-consistent data set that can be compared with
the model results. The observations span some thirty years, and the
spectra were taken with a variety of instruments and reduction
techniques, at different resolutions and with different apertures. The
UV lines were measured with an aperture of $18\arcsec \oslash$ (Kriss
et al. \cite{Kriss}). The apertures of the optical and near-IR spectra
(in those papers where quoted), range from
$1^{{\prime\prime}}\times3^{{\prime\prime}}$ (Boksenberg et
al. \cite{BSAFPS}) to $8\arcsec \oslash$ (Anderson
\cite{Anderson}). The NIR spectra were taken with long slits, ranging from
$2\arcsec\times10\arcsec$ (Thompson \cite{Thompson}) to $3\arcsec\times72\arcsec$
(Osterbrock, Shaw \& Veilleux \cite{OSV}). The apertures of the SWS
($14\arcsec\times20\arcsec$ to $20\arcsec\times33\arcsec$), and LWS ($80\arcsec
\oslash$) are much larger and cover the entire nucleus (See Figure 1 in Sturm et al.
\cite{Sturm}). Since the EELR emission in NGC\,4151 is very extended and
inhomogeneous, these aperture differences cannot be corrected by simply
normalizing the fluxes to the aperture size. Here we attempt no such
correction. This can be justified by noting that Yoshida \& Ohtani
(\cite{YO}) find that 83\% and 78\% of the $\bOIIIbB$ and $\bOIIb$ emission
in the central $28\arcsec\times10\arcsec$ of the nucleus, respectively, is
emitted from the NLR in a $4.3\arcsec\times3.9\arcsec$ region centered on
the continuum source. We expect that the higher ionization species will be
even more centrally concentrated and conclude that the overall error in the
line flux measurement that is introduced by the aperture differences is
smaller than 20\% for all but the smallest apertures.

Broad / narrow component decomposition, variability and absolute flux
normalization are three other inter-related problems. The narrow
component was separated only in the strongest broad lines, while the
quoted values of the weaker permitted lines include both
components. Profile decomposition is an uncertain procedure even under
ideal conditions. In the case of NGC\,4151, it is further complicated
by its high variability. In particular, the 1984 low state of
NGC\,4151 occurred at the middle of the epoch covered by the
observations cited above. Most of the optical lines are given only
relative to the narrow or full $\Hb$ line. The absolute flux value is
given only in a few cases, so that the inter-calibration of these
measurements is uncertain. Finally, in most cases no measurement
errors are given.

Our choice of the lines to include in the modeling is guided by four
criteria. First, we want to consider only lines that are primarly formed by
photoionization by the extreme-ultraviolet and soft X-ray continuum of the
central source. We exclude all lines emitted from ions with $\Eion <
13.6$\,eV, since these can be easily ionized by other processes. In
particular, this criterion leads to the exclusion of the far-IR O\,{\sc
i}\,$\lambda 63.2 \mu$m, O\,{\sc i}\,$\lambda 145.5\mu$m and C\,{\sc
ii}\,$\lambda 157.7\mu$m lines, which are produced mainly in neutral
photon-dominated regions (PDRs) (Tielens \& Hollenbach \cite{TH}; Sternberg
\& Dalgarno \cite{SD}) and in X-ray dissociation regions (Maloney,
Hollenbach \& Tielens \cite{MHT}). Second, we want to consider only lines
with reliable flux measurements. This leads to the exclusion of all optical
lines taken with spectral resolution worse than 20\AA\/, those with fluxes
lower than 1/4 of the narrow $\Hb$ flux, or permitted lines whose narrow
component was not decomposed from the broad one. The HUT narrow UV lines
(Kriss et al. \cite{Kriss}) are somewhat problematic in this respect, since
their FWHM is 1.5 to 3 times larger than that of the NLR forbidden lines
(c.f. Paper I). Although this may indicate a problem in their decomposition
we choose to retain these measurements because of the importance of UV data
for constraining the reddening. We exclude the two weakest ISO-SWS lines,
$\bSiIXbA$ and $\bMgIVb$, which have a low S/N. Third, we require that the
line measurements not be too inconsistent with the large SWS and LWS
apertures. This criterion leads to the exclusion of all optical lines taken
with apertures whose smaller dimension is less than $3\arcsec$, which
unfortunately include the important He\,{\sc ii}\,$\lambda4686$ line. The
$\Eion < 13.6$\,eV criterion also helps in this respect by decreasing the
contamination of EELR line emission. Fourth, we need to take into account
the limitations of the photoionization models. This leads to the exclusion
of the $\bFeXb$\AA\/ and $\bFeXIb$\AA\/ lines, whose collision strengths
values are highly uncertain.  The final, much shortened line list is given
in Table~\ref{t:flux}. Whenever more than one measurement of the line
exists, we quote the average flux and use the rms scatter as an error
estimate. We list also some of the lines and upper limits that are not used
in the model fits, so that their consistency with the model results can be
checked.

\section{The photoionization models}
\label{s:models}
\subsection{The SED / cloud models}
\label{s:SEDcloud}

Two basic ingredients have to be specified for calculating the
photoionization models: the gas, or `cloud' properties and the ionizing
SED. The main objective of this work is to reconstruct the ionizing SED of
NGC\,4151 as model-independently as possible. However, cloud models must be
adopted in the analysis, and these introduce some uncertainties in the
reconstructed SED. The cloud models we consider here (Figure~\ref{f:geom})
are probably too simple to fully reproduce the rich observed nuclear
structure, but they allow us to assess the robustness of the reconstructed
SED to the uncertainties of modeling the gas. 

In the cloud models that are described below, we distinguish between
compact and clumpy gas models. In a compact gas model the entire cloud
volume is filled with gas, that is, its filling factor $F$ is unity. In a
clumpy gas model, $F<1$, and the cloud can be described as composed of many
`cloudlets', or alternatively as having a porous, sponge-like
structure. Only the {\em total} column density of the cloud is specified.
The photoionization calculations assume infinitesimal cloudlets (or holes)
that are randomly and uniformly distributed in the cloud volume. In this
limit, $F$ only enters the photoionization calculations by modifying the
volume emissivity and the optical depths (Ferland \cite{Ferland}). Although
the cloud as a whole may be optically thick, the cloudlets can be thought
of as a system of optically thin clouds that filter, but do not entirely
absorb, the ionizing radiation, and then transmit it to cloudlets further
to the back of the cloud. This picture is physically relevant as long as
the illuminated cross-section of the cloud is not too small relative to its
depth, since it is unlikely that a thin filament of cloudlets can remain
perfectly aligned so as to have the specified total column density. A
clumpy gas distribution `stretches' a given column density over a larger
physical distance relative to that of a compact cloud. Therefore, the
geometrical dilution of the radiation is larger, and hence the ionization
structure of the gas and its emission line spectrum are modified.

We consider four types of clouds.
\begin{enumerate} 
\item Constant density, optically thick clumpy clouds (i.e. $F<1$),
distributed on a spherical shell of radius $r$ centered on the
continuum source. 
\item Constant density, optically thick compact clouds (i.e. $F=1$),
similarly distributed.
\item Optically thick compact clouds, similarly distributed, with a density
gradient that increases linearly into the cloud (i.e. away from the
illuminated surface),
\begin{equation} 
n(d) = n_0(1+d/r_0)\,,
\end{equation} 
where $n_0$ is the gas density at the irradiated surface, $d$ is the
distance into the cloud and $r_0^{-1}$ the gradient. Such models may
approximate clouds with evaporating irradiated surfaces (Binette,
Wilson \& Storchi-Bergmann \cite{BWS}) or a density stratification due
to radiation pressure (Binette et al. \cite{BWRS}; Binette
\cite{Binette}).
\item  Two components of constant density, optically thick compact clouds,
distributed on two spherical shells of radii $r_1 < r_2$, where it is
assumed that the SED seen by component 2 is the same as that seen by
component 1, with no obscuration but with geometrical dilution.
\end{enumerate}  
Realistically, the clouds are likely to have a distribution of
densities and positions relative to the central continuum
source. However, these simple models may be justified by noting that
the large aperture of the SWS (and to a lesser degree the smaller
apertures of the optical and NIR spectrometers) integrates over this
population.  The validity of these models then rests on the assumption
that the population average can be approximated by a single,
representative cloud type.

We model here only ionization bounded (IB) clouds (i.e. optically thick
clouds). Matter bounded (MB) cloud models (i.e. optically thin clouds) have
been proposed in response to discrepancies between the observations and the
predictions of NLR / EELR single component models, which assumed a simple
power law SED (Viegas \& Prieto
\cite{VP}; Binette et al.  \cite{BWS}). Those models typically
under-predict the high ionization lines, under-predict the electron
temperature and under-predict the scatter in the ${\rm He}\,{\sc
ii}\,\lambda4686/\Hb$ ratio among Seyfert 2 galaxies. An additional
MB component comes at the price of three additional free parameters:
the column density of the matter bounded clouds, their fraction
relative to the IB clouds and the fraction of IB clouds that see the
filtered radiation from the back of the MB clouds relative to those
that see the unfiltered radiation. In view of the considerable
complexity introduced by these additional free parameters, and
because, as will be shown below, the high ionization lines are well
modeled by the IB clouds while the gas temperature may not, in fact,
be problematic in NGC\,4151 (Sect.~\ref{s:n4151-gas}), we do not
consider these types of models here.

   In addition to the cloud geometry and density structure, it is necessary
to specify the ionization parameter, which for isotropic emission is
defined as
\begin{equation} 
U = \frac{Q_{\rm ion}}{4\pi r^2 n_0 c}\,, 
\end{equation} 
where $Q_{\rm ion}$ is the ionizing photon emission rate (s$^{-1}$)
and $c$ is the speed of light.  It is also necessary to specify the
cloud column density $N_c$, which we assume to be effectively
infinite, the element abundances, which we assume to be solar, and the
amount of dust, which we assume to be negligible.

The ionizing SED model we use here is based on the observed
multi-wavelength SED of the 1993 campaign (Edelson et al. \cite{Edelson}).
In using it, we are making the assumption that this SED can be used to
approximate the time averaged SED over time scales comparable to the light
crossing time of the NLR, which is of the order of 1000\,yr.

The enumeration on the SED assumes a four-segment broken power-law
(Figure \ref{f:sed_temp}). The observed SED is adopted longward of 1 Ryd
and shortward of 50 keV. In between these two limits, the enumeration
proceeds by choosing all possible combinations of the three luminosity
densities (erg\,s$^{-1}$\,Hz$^{-1}$) at 4, 8, and 30 Ryd ($L_4$, $L_8$
and $L_{30}$, respectively) and connecting them by straight lines in
the $\log E$--$\log L_\nu$ plane. This yields a total of
$13\times13\times3 = 507$ possible combinations. The break energies of
1 and 4 Ryd were chosen because they correspond to the H and He\,{\sc
ii} ionization edges. The break energy of 30 Ryd roughly brackets the
highest ionization potential of the observed lines. The break energy
of 8 Ryd is arbitrary and is intended to allow additional flexibility
in the SED shape. The quantity $\log L_{30}$ varies between 25.4 and
26.0 in increments of 0.3. The lower limit for the $\log L_{30}$ range
is obtained by extrapolating down to 30 Ryd a power-law using
observations at 50 to 80 keV (longward of the power-law break at
$\sim90$\,keV) and at $\sim 4.1$\,keV (shortward of the absorption
feature but still longward of the X-rays Fe emission lines). The upper
limit on $L_{30}$ allows for possible errors in the determination of
the continuum at $\sim 5$\,keV. The values of $\log L_4$ and $\log L_8$
vary between 25.4 and 27.8 in increments of 0.2, thereby reaching
values as high as that at 1 Ryd and as low as that at 30 Ryd. This
allows the enumeration on the SED to explore the possibility of a very
prominent bump, as well as that of a steeply falling continuum in the
1-30 Ryd range.

The photoionization calculations were carried out using the numerical
photoionization code ION97, the 1997 version of the code ION described in
Netzer (\cite{Netzer96}).

\subsection{The fit procedure}
\label{s:fit}
   
The very large parameter space we investigate in this work makes it
necessary to adopt a goodness-of-fit score to rank the models. There are,
however, some difficulties in applying standard methods to the problem at
hand. Both the measurement error estimates and those of the model are
highly uncertain, and are probably non-Gaussian and dominated by
systematic errors. Therefore, the commonly used $\chi^2$ score, which
relies on the accuracy of the errors estimates, can be highly
misleading. Furthermore, the model lines and the observed lines are not
directly comparable, since the model calculations yield the line luminosity
per unit area on the cloud's cross-section, whereas the observations are of
the line flux on earth.  A standard procedure for comparing the two is to
normalize all the line fluxes relative to the flux of some strong reference
line, such as $\Hb$, which is roughly proportional to the total ionizing
luminosity. However, in the case of NGC\,4151, this is problematic because
the narrow $\Hb$ flux measurement is very uncertain due to the difficulty
of decomposing the narrow and broad components. The observed forbidden
optical and IR lines are also not suited for this purpose because of the
large measurement errors. An error in the flux of the chosen reference line will
bias all the line ratios, and because of the highly non-linear behavior of
the photoionization calculations, this is likely to have complicated
effects on the best-fit SED. Moreover, the freedom of choosing the reference
line adds ambiguity to the results, since there is no compelling reason for
preferring one reference line over another.

Here we formulate and adopt a different fit procedure that does not
rely on the accuracy of the measurement error estimates and treats all
the lines on an equal footing. We assume that for the correct model
and for ideal, error-free observations, the observed and model line
fluxes are related by
\begin{equation}
\exp(\kl\EBV)\fl = \alpha \ml\,,
\end{equation}
where $\fl$ is the observed flux in line $\ell$, $\ml$ is the model
luminosity per unit surface area of the irradiated face of the cloud,
and the exponent is the dereddening factor with $\kl$ given by the
extinction curve at the line's rest wavelength. The parameter $\alpha$
is a proportionality factor common to all the lines. It has a simple
geometrical interpretation in the framework of the cloud models we
consider here. These models assume an isotropic line emission from
clouds distributed on the surface of a sphere of radius $r$, which is
centered on the continuum source. In this case
\begin{equation}
\exp(\kl\EBV)\fl = C 4\pi r^2 \ml / 4\pi D^2\,,
\end{equation}
where the covering factor, $C$, is the fraction of the spherical surface
covered by the clouds and $D$ is the distance of the AGN from the
observer. Therefore,
\begin{equation}
\alpha = C \left(\frac{r}{D}\right)^2\,.
\end{equation}
The distance $r$ and the corresponding angular distance $\theta$ are fixed
by the ionization parameter and the hydrogen gas density,
\begin{equation}
\label{e:r2}
r^2 = \frac{D^2\qion A}{Un_0c}\,,
\end{equation}
\begin{equation}
\label{e:theta}
\theta^2 = \frac{\qion A}{Un_0c}\,,
\end{equation}
where $A$ is the anisotropy parameter of the ionizing luminosity (Sect.
\ref{s:n4151-nucleus}), and $\qion$ is the ionizing photon flux that would
be observed if all the ionizing photons reached the observer. The parameter
$\qion$ is a function of $L_4$, $L_8$ and $L_{30}$, which parameterize the
SED. Equation \ref{e:theta} can be used to estimate $A$ if the other
parameters are known independently. Here, this information is unavailable,
and $A$ remains a free parameter. The covering factor is related to
$\alpha$ by
\begin{equation}
\label{e:cf}
C = \frac{\alpha}{\theta^2}\,.
\end{equation}

The depth of the ionized fraction of the gas distribution, $d_{\rm ion}$,
which is calculated by the photoionization code, can be used to estimate
the angular extent of the NLR, $\tion$,
\begin{equation}
\label{e:tion}
\theta_{\rm ion} = d_{\rm ion}/D = d_{\rm ion}h/cz\,.
\end{equation}
$\tion$ corresponds to the angular extent of the Balmer lines
emission. Emission from higher ionization lines is more concentrated
towards the inner part of the NLR.

We note that there exists a scaling relation between $A$ and $F=F_1h$
(observations fix $F$ only up to a factor of $h$) that allows models, which
differ only in these parameters, to have identical best-fit line fluxes and
SEDs. This can happen when both $U$ and the geometrical dilution of the
ionizing flux, 
\begin{equation}
g(d) = (1+d/r)^{-2}\,,
\label{e:geom_dilute}
\end{equation}
are unaffected by the changed parameters. Since $d$ scales as $1/F$,
$g$ remains constant if $r$ is also scaled as $1/F$, which in turn
implies that $U$ should scale as $A F_1^2$. Therefore, if $A$ scales
as $1/F_1^2$, the line ratios remain constant, while $\theta$ and
$\tion$ scale as $A^{1/2}$ and $C$ as $1/A$.

The search for the best combination of cloud model and SED proceeds by
enumerating over various possible cloud models, whose input parameters are
$U$, $n_0(d)$, $F$, $A$, the number of gas components and their elemental
abundances, and the `boundary conditions' of the SED template, $L_1$ and
$L_{4134}$. In addition, $h$ is required for calculating $r$ and $\tion$
and also for estimating the input value of $F$ from the observations.  For
each cloud model, we then enumerate over $L_4$, $L_8$ and $L_{30}$, and use
a `goodness-of-fit' score, which is described below, to find the values of
$L_4$, $L_8$, $L_{30}$, $\alpha$ and $\EBV$ that best fit the observed
line fluxes. The observational constraints on $\EBV$ and the derived
quantities $C$, $\theta$ and $\tion$ can then provide additional checks on
the best fit values of these parameters.

Lacking reliable measurement and model error estimates, a natural measure
of the match or mismatch between model and observations is simply the line
ratios, $\ml/\fl$. For this reason, photoionization models are commonly
judged by their ability to fit the data to within a factor of $S$, on
average, where $S<2$ is considered to be a reasonable fit. We formalize
this intuitive concept of a logarithmic scale by defining a score function
\begin{equation} 
\lgs = \frac{1}{n}\sum_{\ell=1}^n\left(\log\frac{\alpha \ml}
{\exp(\kl\EBV)\fl}\right)^2\,,
\end{equation}
where $n$ is the number of lines used in the fit.  The $\lgs$ score is an
explicit function of $\alpha$ and $\EBV$ and an implicit function of $L_4$,
$L_8$, $L_{30}$ and the cloud model through the model line fluxes $\ml$. An
average fit factor $S$, in the rms sense, is defined as
\begin{equation} 
S = \exp\left(\sqrt{\lgs}\right)\,,
\end{equation} 
with the worst fitting line off by a factor of
\begin{equation}
\max S_\ell = \max_\ell \exp\left(\left|\log \frac{\alpha^0 m_\ell^0}{\exp(\kl
E^0_{\sc b-v})\fl}\right|\right)\,.
\end{equation} 

It is straightforward to show that for a given set of model fluxes,
$\{\ml\}$, $\lgs$ is minimized when $\alpha$ and \EBV take the values
\begin{equation}        
\log\alpha^0 =
\frac{\langle k\rangle\langle k \log m/f\rangle-
\langle k^2\rangle\langle\log m/f\rangle}
{\langle(k-\langle k\rangle)^2\rangle}\,,
\end{equation}
and
\begin{equation} 
E_{\sc b-v}^0 =
\frac{\langle(k-\langle k\rangle)(\log m/f-\langle\log m/f\rangle)\rangle}
{\langle(k-\langle k\rangle)^2\rangle}\,,
\end{equation} 
where the notation $\langle\ldots\rangle$ designates the average of the
bracketed quantities over the $n$ lines, and it is assumed that not all the
\kl are equal. $E_{\sc b-v}^0$ can formally take negative, non-physical
values. In such cases we set it to zero, and re-minimize \lgs as a function
of $\alpha$ only, in which case
\begin{equation}
\log\alpha^0 = -\langle\log m/f\rangle\,.	 
\end{equation} 
The search for the minimum of \lgs in $L_4$, $L_8$, $L_{30}$, \EBV and
$\alpha$ proceeds by first calculating numerically $\{\ml\}$ for each point
in ($L_4$, $L_8$, $L_{30}$) space and then analytically evaluating \lgs at
each point using $\alpha^0$ and $E^0_{\sc b-v}$.

   \lgs is easily generalized to the case of two cloud components,
\begin{equation} 
\lgs = \frac{1}{n}\sum_{\ell}\left(\log\frac{
\alpha (w_1 m_{\ell1}+w_2 m_{\ell2})}
{\exp(\kl\EBV)\fl}\right)^2\,,
\end{equation}
where $w_1+w_2 = 1$ are the mixture weights of the two components,
which emit a line luminosity $m_{\ell1}$, $m_{\ell2}$ per unit surface
area, respectively.  The covering factor of each component is given by
Eq.~(\ref{e:cf}) with $\alpha w_1$ or $\alpha w_2$ in place of
$\alpha$. The \lgs score is minimized by numerically enumerating on
possible values of $w_1$, calculating $\ml = w_1 m_{\ell1}+w_2
m_{\ell2}$ and then proceeding as in the case of a single cloud
component.

The errors on the best-fit parameters of a given cloud model, which are
introduced by the measurement errors in the line fluxes, are estimated by
Monte Carlo simulations (see e.g. Press et al. \cite{PTVF}). A set of
simulated observations $\{f^\prime_\ell\}$ are drawn using the best-fit
parameters and the given measurement error estimates (or a conservative
guess of a 50\% error if such an estimate is unavailable),
\begin{equation}
f^\prime_\ell = \alpha^0 m^0_\ell
                \exp(-k_\ell E^0_{\sc b-v})(1+\epsilon)\,,
\end{equation} 
where $\epsilon$ is a Gaussian deviate of zero mean and standard deviation
equal to the quoted fractional measurement error of the line. The
minimization procedure is then repeated with the given cloud model for each
simulated set, and the best-fit parameters are recorded. The 99.9\%
confidence limits, which are quoted below, lie between the minimal and
maximal values that the parameters take on the contour of constant $S$ that
encloses 99.9\% of the simulated results. The errors are given for each
parameter separately and are {\em not} statistically independent. Note that
this is the only aspect of the modeling where the measurement error
estimates affect the results. Note also that the confidence limits on the
SED are confined to the range covered by the SED enumeration (see
Figure~\ref{f:sed_temp}). The quoted confidence interval may in fact be
smaller than the true one in places where it extends right up to the edge
of the enumerated region. It should also be emphasized that the confidence
limits are conditional, in the sense that they are calculated under the
assumption of a given gas model and SED template. We present below errors
only on the parameters $L_4$, $L_8$, $L_{30}$, $\EBV$, $\theta$, $\tion$
and $C$. The Monte Carlo procedure can be easily generalized so as to
obtain confidence limits on all the free parameters, including those of the
gas model, $U$, $F$, $n_0$ and $r_0$. However, because we test only a small
number of values for these parameters, this is not attempted here.

The weighting strategy is a major concern in any fit procedure, including
this one. The $\lgs$ score assigns equal weights to the different lines. In
doing so, we are ignoring the fact that some lines carry more physical
information than others, and that the unequal spacing of the lines in
$\Eion$ over-emphasizes some energy ranges in the SED at the expense of
others. Nevertheless, because it is unclear how to deal in a satisfactory
and general way with these problems, and because of the simplicity and
elegance of the $\lgs$ fit, we choose to use the simplest option of equal
weights. As will be shown in Sect.~\ref{s:discuss}, we test the sensitivity
of the best-fit SED to the weighting by dropping a subset of the lines from
the fit (i.e. setting their weight to zero) and re-fitting. We find that
the best-fit SED is insensitive to these changes.

   A combined SED and cloud model is considered successful if it fulfills the
following criteria.
\begin{enumerate}
\item
   The model fits the lines to within a factor of 2, on average ($S < 2$).
\item 
   The worst fitting line is off by no more than a factor of 3 ($\max
   S_\ell< 3$). 
\item 
   The covering factor is less than 0.25, which corresponds to a
   bi-cone with an opening angle of $\sim 80^\circ$ ($C \la 0.25$).
\item 
   The angular distance of the illuminated face is consistent with the
   inner radius of the NLR ($\theta \la 0.5\arcsec$).
\item
   The angular extent of the line emitting gas is consistent with the size
   of the NLR ($\tion \sim 5\arcsec$).
\item 
   The extinction is small ($\EBV < 0.15$).
\end{enumerate} 
We also attempt to detect systematic trends in the residuals, or remaining
discrepancies, of the best-fit model. This is done by calculating the
correlation between the best-fit ratios, $\alpha^0 m_\ell^0/\exp(k_\ell
E_{\sc b-v}^0)f_\ell$, and the line properties $\lambda_0$, $\Eion$ and the
element's tendency for depletion. These test, respectively, for problems
with the assumed extinction law, problems with the assumed SED
parameterization, and problems with the elemental abundances (assumed
solar), which may be depleted by dust. In order to minimize the sensitivity
of these tests to numeric uncertainties and to assumptions on the nature of
the correlations (e.g. linear in the logarithm of the quantity vs. linear
in the quantity itself), we use the non-parametric Kendall's $\tau$
correlation coefficient (see e.g. Press et al. \cite{PTVF}). Using this
estimator, the possibility of depletion is simply tested by correlating the
best-fit ratios (after summing line fluxes from the same element) with the
depleted abundances, which we take to be those of the Galactic interstellar
medium given by Spitzer (\cite{Spitzer}). A negative correlation is
expected if the true abundances are depleted. The residuals of the correct
model should not display any such correlations.

\section{Results}   
\label{s:results}   

The cloud models that we investigate in this work (Table~\ref{t:score})
include clumpy and compact single component models with constant density,
compact single component models with a density gradient, and two component
models that result from mixing pairs of compact, single component constant
density models.  The ionization parameters lie in the range $U=0.005$ to
$0.4$ and the filling factors in the range $10^{-3}h$ to $1$.  We assume
density profiles of the form $n (r) = n_0(1+d/r_0)$, where
$n_0=1000$\,cm$^{-3}$ and $r_0 = 3\,10^{17}$\,cm or $r_0 = \infty$, the
latter corresponding to the constant density case. We assume throughout
$h=0.65$, isotropic source emission ($A=1$), solar abundances, no dust, and
use a column density that is large enough to absorb all the ionizing
radiation. For each of these cloud models, we enumerate on the SED template
(Figure~\ref{f:sed_temp}), calculate the resulting line fluxes with the
photoionization code and perform the fit procedure. The extinction curve
used for dereddening is based on Seaton (\cite{Seaton}) for the UV and
optical and Lutz et al. (\cite{Lutz2}) for the IR. Because of the
uncertainty involved in inter-calibrating the optical and ISO flux
measurements, we repeat the fit procedure twice, once using only the ISO
line list (`ISO fit') and once with the full ISO and optical line list
(`full fit'). This also allows us to assess the sensitivity of the equal
weighting scheme that underlies the $\lgs$ score.

The range of ionization parameters was chosen after spot-checks indicated
that lower $U$ models severely under-produce the high ionization lines
emission, while higher $U$ models over-produce them. The filling factor
$F=0.065$ corresponds to that used to model the Circinus SED by Moorwood
et al. \cite{Moorwood}, and that of $6.5\,10^{-4}$ is the value inferred
for the low excitation clouds of the NLR in NGC\,4151 (Yoshida \& Ohtani
\cite{YO}); Robinson et al. \cite{Robinson}). We also made some
spot-checks with models of higher and lower densities ($n_0 = 300$ and
$3000$\,cm$^{-3}$) and larger and smaller density gradients ($r_0 = 10^{17}$
and $10^{18}$\,cm). These modified models gave qualitatively similar
results to the models included in the grid, and will therefore not be
further discussed.

\subsection{Single component models}

The best-fit scores of the single component models are listed in
Table~\ref{t:score}. As expected, the ISO fits, which are less
constrained, have better $S$ scores than the full fits. Of the six fit
criteria ($S$, $\maxs$, $C$, $\theta$, $\tion$ and $\EBV$), $\EBV$ is
the least restrictive (Some high $U$ models have ISO fits with very
large $\EBV$ values, but these have very large errors). The covering
factor $C$ points to models with low ionization parameter or small
filling factor. The angular distance $\theta$ is uncomfortably large
for the lowest $U$ models, $S$ and $\maxs$ disfavor models with the
lowest ionization parameters and constant density models with high
filling factors. The most restrictive criterion turns out to be
$\tion$. Only models with the lowest filling factor have the large
observed NLR size. 

   Taking into account all the criteria, the best model, for both the
ISO and full fits, is that with $U=0.025$, $F=6.5\,10^{-4}$ and a
constant density. This particular choice of $U$ is somewhat arbitrary,
as the higher $U$ models in the $F=6.5\,10^{-4}$ sequence fit almost
equally well (see Sect.~\ref{s:discuss}). The Balmer lines NLR emission
in this model extends between $\sim 0.5\arcsec$ to $\sim 5\arcsec$.
The $\bOIIIbB$ emission is more centrally concentrated, and extends
only up to $\sim 3\arcsec$. This is consistent with the images of
Hutchings et al. (\cite{Hutchings}). The covering factor of this
model, $C=0.24$, implies that the gas fills almost the entire NLR
bi-cone. With such a large cloud cross-section, the clumpy gas
description is physically reasonable (Sect.~\ref{s:SEDcloud}).

The properties of the best-fit model are summarized in Table~\ref{t:fit},
the best fitting SEDs are shown in Figure~\ref{f:best_SED} and the best-fit
line ratios are shown in Figures \ref{f:ratio_ISO} and
\ref{f:ratio_full}. Apart for a somewhat large $\maxs$ in the full fit,
both fits are successful in the sense defined in
Sect.~\ref{s:fit}. The results of the two fit procedures are very
similar. In particular, in both the SED falls steeply beyond 1 Ryd
towards 4 Ryd and then rises again at around 8 Ryd.  As can be
expected, the ISO fit does not constrain $\EBV$ very well and also
does not reproduce the optical lines as well as the full fit. Although
the confidence limit on $\EBV$ are improved in the full fit, they are
still quite large. This probably reflects the fact that the fit is
still dominated by the IR lines, and that the optical reddening
indicators give conflicting estimates (Sect.~\ref{s:n4151-gas}). The
best fit $\EBV = 0.03$ is consistent with the lower range of these
reddening estimates. In both fits, even the lines that were excluded
from the $\lgs$ score are reproduced reasonably well, including the
He\,{\sc ii}\,$\lambda4686$ line, for which wide aperture measurements
are unavailable.

\subsection{Two component models}

The fitting procedure of the two component models treats the mixture
weight, $w_1$, as an additional free parameter of the fit. We find that in
most cases, the best-fit two component model was one with the trivial
mixture weight of $w_1=0$ or $1$, meaning that the fit of the better model
of the two could not be improved by any amount of mixing with the other
model. In those cases where a non-trivial mixture did improve the fit, the
score was still worse than that of the best fitting single component model,
for both the full and ISO fits. We therefore conclude that there is no
compelling reason to prefer the mixed two-component constant density models
over the simpler single component models.

\subsection{Dependence on the UV template SED}
\label{s:UV}
The template SED has two points fixed by the observations, one at 1 Ryd and
one at 56.2\,keV (4134 Ryd). The hard photons at 56.2\,keV are well beyond
the ionization potentials of the lines and affect only the low ionization
lines by maintaining partially ionized regions at the back of the
clouds. Therefore, the model line fluxes should not depend strongly on the
exact SED value at 56.2\,keV. This is not necessarily the case for the
fixed point at 1 Ryd, as these photons interact very strongly with the gas.
There are two sources of uncertainty in the value adopted for the SED at 1
Ryd. One is the strong variability of NGC\,4151, which may result in a big
difference between the template SED and the effective, time-averaged SED
that is relevant for the NLR. The other is the accuracy of dereddening
correction that was used to derive the intrinsic SED from the observed one.

In order to check the sensitivity of the results to the shape of the UV
SED, we repeated the calculations for the clumpy, $F=6.5\,10^{-4}$, model,
this time varying also the flux at 1 Ryd. Instead of the fixed value of
$\log L_1 = 27.88$, we enumerated on the values $\log L_1=27.8$, 28.0 and
28.2. These correspond to the dereddened IUE flux at 1806\AA\/ with
$\EBV=0.03$, $0.08$ and $0.11$, respectively, or to any combined effect of
reddening and variability of order $\Delta\log L_1 = 0.4$. For reasons of
computational economy, we enumerated only on every second point in $L_4$
and $L_8$. The best-fit results of these models and those of the fixed 1
Ryd luminosity models are almost identical. We therefore conclude that
uncertainties of the order of 1 mag in the determination of the luminosity
at the Lyman limit do not have a significant effect on the best-fit
results.

\section{Discussion}
\label{s:discuss} 
\subsection{Properties of the best fit model}

The best-fit model succeeds in reproducing the observed line fluxes to
within a factor of 2, on average. It is also consistent with low
extinction and with the geometrical constraints on $\theta$, $C$ and
$\tion$. We find no indication of a Big Blue Bump rising beyond the
Lyman limit (c.f. Figure~\ref{f:sed+bump}). On the contrary, the SED
declines steeply from the Lyman limit towards 4 Ryd and then rises
sharply and peaks at around 8 Ryd.  The exact shape of the SED varies
among the various successful models, but generally, we get very
similar results from the different cloud models we considered here.

The best-fit model is a single component, clumpy cloud with constant
density. We find no compelling evidence pointing towards multi-component
clouds, non-solar elemental abundances or significant continuum anisotropy.
The $A/F$ scaling relation (Sect.~\ref{s:fit}) cannot be applied in this
case without increasing $\theta$ and $\tion$ beyond their observed values
of $\la 0.5\arcsec$ and $\sim 5\arcsec$, respectively. Note, however, that
$A$ is not well constrained because of the uncertainty in $\qion$ (cf
Eq.~\ref{e:theta}) that follows from the uncertainty in $L_1$
(section~\ref{s:UV}). A clumpy gas distribution enhances the geometrical
dilution of the ionizing radiation relative to a compact distribution by
`stretching' a given gas column density over a larger physical distance. It
appears that geometrical dilution plays an important role in establishing
the ionization structure that is required for reproducing the observed line
ratios. This can be seen by noting that models with a density gradient
(e.g. $U=0.01$, $\log r_0 = 17.5$) also reproduce the line ratios very
well, and result in the same best-fit SED. In such models, geometrical
dilution is qualitatively mimicked by the increase in gas density with
depth into the cloud. However, such models fail to reproduce the large
extent of the NLR since only the thin illuminated skin of the cloud emits
lines. It is possible that an extended system of such clouds with suitably
adjusted ionization parameters may be consistent with the observations as
well, and may be natural in the context of radiatively accelerated clouds
(Binette
\cite{Binette}). However, the dynamics of the NLR are outside the
scope of this work, and in view of the success of simpler models, this
possibility was not investigated here.

The insensitivity of the low-$F$ results to the value of $U$
(Sect.~\ref{s:results}) is also related to the geometrical dilution. The
volume averaged ionization parameter $\bar{U}$ (taking into account
only geometrical dilution but ignoring true absorption) is
\begin{eqnarray}
\bar{U} & = & U\left.\int_{r_1}^{r_2}r^2 g(r)\,{\rm d}r \right/
                     \int_{r_1}^{r_2}r^2\,{\rm d}r \nonumber\\
        & = &  3U\left/\left[1+r_2/r_1+(r_2/r_1)^2\right]\right. \nonumber\\
        & \stackrel{\textstyle \longrightarrow}{_{r_2\gg r_1}} & 3g(r_2)U\,,
\end{eqnarray}
where $r_1$ and $r_2$ are the inner and outer radius of the NLR,
respectively, and $g(r) = (r_1/r)^2$ is the geometrical dilution (c.f.
Eq.~\ref{e:geom_dilute}). When the geometrical dilution is significant,
$\bar{U}$ can be much smaller than $U$, which is defined at the
irradiated face of the cloud, since the volume average is dominated by
the outer shells. The inner and outer radii of the NLR are free
parameters in our fit procedure. Inspection of the $F=6.5\,10^{-4}$,
$U\ge 0.01$ full fit results in Table~\ref{t:score} shows that the
best fit value of $\tion/\theta$ varies with $U$ in such a way that
$g(r_2) \propto U^{-1}$ approximately, and therefore $\bar{U}
\sim 10^{-3} \ll U$ irrespective of $U$. In addition, the best fit
SEDs of all these models are almost identical to that of the $U=0.025$
model (Figure~\ref{f:best_SED}). Thus, while the emission line data
cannot precisely constrain $U$, $\theta$ or $\tion$ individually, they
clearly favor models with a low $\bar{U}$, a low $F$ and a very hard
SED.

The filling factor of the best-fit model, $F=6.5\,10^{-4}$, was deduced
from observations of the low excitation NLR lines. The high ionization
lines of the best-fit model are produced closer to the central source,
while the lower ionization lines are produced deeper in the cloud in less
ionized regions. The fact that this model reproduces both the high and low
excitation lines supports a simple picture that the clumps emitting the
coronal lines and those emitting the optical narrow lines form one
continuous and homogeneous system.

The most significant result of this work is that we find no evidence for a
Big Blue Bump just beyond the Lyman limit of NGC\,4151. Such a bump is too
soft to be consistent with the flux ratios of the high and low ionization
lines. This is demonstrated in Figure~\ref{f:gray}, which shows the model
to data ratio ($\alpha\ml/\exp(\kl\EBV)\fl$) of the $\bArIIIb$ line
($\Eion=27.6$\,eV) and $\bSiIXbB$ line ($\Eion=303.2$\,eV) as function of
$L_4$ and $L_8$ for the best-fit model. Both lines are very well modeled
without a Big Blue Bump (Figure~\ref{f:ratio_full}). The right hand side of
the gray scale plots corresponds to high luminosity at 4 Ryd, and the lower
right hand side corner to the case of an extreme Big Blue Bump, which peaks
at around 4 Ryd ($\sim 50$\,eV) and then falls off steeply. Such a bump
clearly over-produces the low ionization line and under-produces the high
ionization line. 

A single power-law is also not favored by the observations. As an example,
we calculated the fit for a model similar to the best fit model, but
with $\log L_4= 26.6$, $\log L_8 = 26.4$ and $\log L_{30} = 25.7$, which is
the SED that most closely resembles a single power-law and is still
contained within the 99.9\% confidence interval shown in
Figure~\ref{f:best_SED}. This SED obtains a fit score of $S = 2.8$,
under-predicts the $\bMgVIIIb$ line by a factor of 8.5, and under-predicts
the other high $\Eion$ lines as well. The fit of this model to the observed
NLR geometry is also clearly worse than that of the best fit model. The
softer SED results in a smaller ionization depth, $\tion = 3.6\arcsec$, a
higher covering factor, $C = 0.35$, and a higher photon flux $\qion$, so
that $\theta = 0.83\arcsec$ (Eq.~\ref{e:theta}).

\subsection{Comparison with the Circinus SED}

It is instructive to discuss the SED of NGC\,4151 by comparing it to
the SED of the Seyfert 2 Circinus galaxy. Unlike NGC\,4151, the
reconstructed SED of the Circinus galaxy does appear to have a Big
Blue Bump (Moorwood et al. \cite{Moorwood}). As a check on the
Circinus results and a test of the $\lgs$ method, we applied the
fitting procedure described here to the observed optical and IR line
fluxes of Circinus (Oliva et al. \cite{Oliva}; Moorwood et
al. \cite{Moorwood}). We assume, as in Moorwood et
al. (\cite{Moorwood}), a single component constant density clumpy gas
distribution of solar abundances with $n_0 = 5000$\,cm$^{-3}$, an
ionization parameter $U=0.45$, a filling factor $F=0.1h$, isotropic
emission ($A=1$) and $h=0.65$.  Figure~\ref{f:best_cir} shows the
best-fit SED for this model.  Only lines with $\Eion>30$\,eV were used
in the fit to avoid contamination from lines excited by hot stars,
which contribute significantly to the nuclear emission (Moorwood \&
Oliva \cite{MO}; Marconi et al. \cite{Marconi}; Moorwood et
al. \cite{Moorwood}). This model fits 17 observed lines with $S=1.7$,
$\maxs = 2.8^{-1}$, $\EBV=2.0$, $C=0.05$, $\theta=0.3\arcsec$,
$\tion=1.9\arcsec$ and $\log Q_{\rm ion}= 53.73$ (for $h=0.65$).
These values are consistent with the observed properties of the
Circinus galaxy (Oliva et al. \cite{Oliva}). A Big Blue Bump is seen
in our best $\lgs$ fit, in qualitative agreement with the Moorwood et
al. (\cite{Moorwood}) result. However, this result cannot be
established with a high confidence level due to the lack of
constraints on the SED below 30\,eV (note that Figure~\ref{f:best_cir}
shows only a 90\% confidence region).  Although the confidence limits
on the best-fit Circinus SED can accommodate a NGC\,4151-like trough,
this is not required by the narrow emission line data. In the
following, we adopt the simplest working hypothesis, namely that the
Circinus SED has the UV bump that is seen in the best-fit SED. We note
that careful separation of the nuclear and stellar contributions is
crucial in studies of this type. When we also include in the fit
star-contaminated lines with $13.6<\Eion<30$\,eV, we find that the
best-fit SED no longer exhibits a bump, but rather falls off sharply
from 1 Ryd to 4 Ryd and then remains flat.

The possibility that the SEDs of Seyfert 1 and Seyfert 2 galaxies are
intrinsically different may have far reaching consequences for unification
schemes of AGNs (Antonucci \cite{Antonucci}). This cannot be reconciled
with the idea that the observational differences between these two AGN
types are due to orientation effects alone. Bearing in mind the large
uncertainties in the reconstruction of the Circinus SED, we make a
tentative attempt to understand the differences between the NGC\,4151 and
Circinus results. We consider first the narrow emission line spectra. A
simple comparison of the NGC\,4151 to Circinus emission line ratio as
function of $\Eion$ (Figure~\ref{f:cir2ngc}) reveals a strong decrease with
$\Eion$. This trend implies that the NLR of NGC\,4151 is not as highly
ionized as that of Circinus. However, our analysis shows that the SED of
NGC\,4151 is as hard or harder than that of Circinus. The mean ionizing
photon energy (between 1 to 30 Ryd) is 3.2\,Ryd (ISO fit) to 4.1\,Ryd (full
fit) for NGC\,4151 as compared to 3.2\,Ryd for Circinus.  This is in the
opposite sense needed to explain the trend illustrated in
Figure~\ref{f:cir2ngc}. It is possible, in principle, that the paucity of
UV photons immediately beyond the Lyman edge in the harder NGC\,4151 SED is
a bottle-neck for the ionization sequence that leads to the high
ionizations levels. However, this is not supported by the trends apparent
in Figure~\ref{f:gray}, which show that modifying the SED by replacing hard
photons by soft photons only further decreases the ionization level. The
geometrical dilution in the NLR of both AGNs is quite similar
($\tion/\theta=6.6$ in Circinus as compared to $\tion/\theta=7.2$ in
NGC\,4151), so that the difference in $\bar{U}$ simply reflect that in $U$,
with $\bar{U}=0.02$ in Circinus being 20 times larger than $\bar{U}=0.001$
in NGC\,4151.

We investigate the relative contribution of these factors to the NLR
ionization by calculating the ratios between the ISO line fluxes in the
best full fit NGC\,4151 model and those in a sequence of four NLR models:
(i) the best fit Circinus model; (ii) a model identical to the best full
fit NGC\,4151 model, apart for having the high Circinus $U$; (iii) a model
identical to the best full fit NGC\,4151 model, apart for having the
Circinus-like bumpy SED shown in Figure~\ref{f:abs_SED}; and (iv) a model
identical to the best full fit NGC\,4151 model, apart for having both the
Circinus $U$ and a bumpy SED. The ratios are plotted in
Figure~\ref{f:ngc_models} as function of $\Eion$. The run of ratio (i) with
$\Eion$ is generally similar but not the same as that based on the observed
line fluxes (Figure~\ref{f:cir2ngc}) because the models don't exactly
reproduce the observations (in particular, the discrepancies in reproducing
the $\bSIVb$ and $\bOIVb$ lines mask the observed trend below
$\sim50$\,eV). We use these model ratios as the standard for comparison
with the other three models. We find that ratio (ii) falls off with $\Eion$
even more sharply than (i). This shows that the large $\bar{U}$ difference
can more than account for the much higher ionization in Circinus. Ratio
(iii) rises with $\Eion$. This demonstrates that all other factors being
equal, the harder SED of NGC\,4151 can ionize the NLR to levels higher than
in Circinus. Ratio (iv) does fall off with $\Eion$, but less than ratio
(ii). We therefore conclude that the $\bar{U}$ difference dominates the
relative line strengths in the spectra of the two AGN.  Although the
relative hardness of the two SEDs affects the line ratios in the opposite
sense, the effect is smaller and it does not reverse the trend.

Next, we attempt to interpret the difference in the SEDs. We begin by
assuming that they are the intrinsic SEDs, i.e. that they directly
reflect the accretion process. There is evidence that the two AGNs
have different radiative efficiencies. Edelson et al. (\cite{Edelson})
estimate that the black hole mass in NGC\,4151 is $4-10\,10^7\,\Mo$,
and that it is emitting at a rate of $L/L_{\sc e}\la 0.01$, where
$L_{\sc e}$ is the Eddington luminosity.  Maiolino et
al. (\cite{Maiolino}) place an upper bound of $4\,10^6\,\Mo$ on the
black hole mass in Circinus, which together with a non-stellar
luminosity estimate of $L\sim10^{10}\,\Lo$ (Moorwood et
al. \cite{Moorwood}), translates into $L/L_{\sc e}> 0.1$. There are
two known families of {\em steady-state} accretion solutions: hot thin
disks (Shakura \& Sunyaev \cite{SS}), and advection-dominated
accretion flows (ADAFs) (Narayan \& Yi \cite{NY}; Narayan et
al. \cite{NKH}). Naked thin disks (i.e. without additional components)
emit thermally and are therefore efficient compared to ADAFs, where
most of the energy is advected into the black hole by the ions, and
only a small fraction of it is emitted through synchrotron radiation
and inverse Compton scattering. The high efficiency of Circinus and
the bump in its SED are naturally explained by thin disk
accretion. However, the reconstructed SED of NGC\,4151 does not fit
our pre-conceived ideas about accretion flow spectra. Neither the high
luminosity of NGC\,4151 nor the sharp features in its SED can be
easily reconciled with ADAF spectra (c.f. Lasota et
al. \cite{Lasota}). Hybrid ADAF / thin disk models such as considered
by Lasota et al. (\cite{Lasota}) can do better in matching the
observed high luminosity, but would re-introduce a significant thermal
bump component. The mismatch between the spectra of these accretion
models and NGC\,4151 may be related to the AGN's extreme
variability. It is conceivable that such high variability can no
longer be described as a small perturbation on top of a steady-state
accretion flow, but is related to some other transient flow geometry.

\subsection{Internally absorbed ionizing continuum}

The alternative assumption is that the reconstructed SED of NGC\,4151 {\em
is not} the intrinsic one, but only the one that photoionizes the NLR. This
can be the case if the NLR is partially shielded from the continuum source
by an absorbing medium (Halpern \& Steiner (1983) suggested that varying
amounts of shielding can naturally explain the Seyfert to LINER sequence of
NLR spectra). The relatively high NLR covering factor of the best fit
model, together with the large broad to narrow $\Hb$ flux ratio, suggest
that this is indeed the case in NGC\,4151. The measured broad to narrow
$\Hb$ ratio varies from 1.5 (DeRobertis \cite{DeRobertis}) during a low
state epoch of the AGN (Penston \& P\'erez \cite{PP}), through $\sim 6$
(Osterbrock \& Koski \cite{OK}), up to $\sim 10$ (Kaspi et
al. \cite{Kaspi}) during a high state epoch\footnote{Based on the low
resolution Wise spectrum of the full $\Hb$ line and the narrow $\Hb$ flux
given in Table~\ref{t:flux}.}  (Kriss et al. \cite{KDZKE}). We adopt a
broad to narrow $\Hb$ flux ratio of 5 as a representative value. The $\Hb$
line emission in the best fit model is well modeled by case B recombination
with $T=10^4$\,K, which is close to the maximal efficiency for $\Hb$
production by recombination per unit covering factor. The covering factor
of the best fit model is $C=0.24$, which implies an unphysical covering
factor for the BLR, $C_{\rm BLR} = C\times5> 1$. It is well known (Netzer
\cite{BNW}) that the uncertainty in calculating the broad Balmer line
intensities is large, due to complicated radiative transfer effects. It is
likely that these effects, combined with collisional excitation of $\Ha$
and $\Hb$, can enhance the $\Hb$ intensity over its recombination value, by
a factor of a few. However, this is unlikely to lower $C_{\rm BLR}$ to an
acceptably low value, especially if the broad to narrow $\Hb$ flux ratio is
in fact as large as $\sim 10$.

A possible solution to this problem is to assume that the BLR gas sees more
ionizing photons than the NLR. For example, the unabsorbed Circinus-like
bumpy SED, shown in Figure~\ref{f:abs_SED}, has 10 times as many ionizing
photons as the best fit SED. In this case, $C_{\rm BLR} = C\times5/10 \sim
0.1$, which is the canonical value for $C_{\rm
BLR}$. Figure~\ref{f:abs_SED} shows that the rise towards 8 Ryd in the best
fit SED is well reproduced by the recovery slope of an intrinsically bumpy
SED absorbed by $\sim 5\,10^{19}$\, cm$^{-2}$ of neutral hydrogen, located
between the BLR and the NLR.  The covering factor of the absorber relative
to that of the NLR, $C_{\rm abs}$, has to be close to unity, but some
leakage is still consistent with the best fit SED.

It is remarkable that such an absorber was detected with completely
independent methods by absorption lines (Kriss et al. \cite{Kriss,KDZKE})
(Sect.~\ref{s:n4151-nucleus}). The absorber's H\,{\sc i} column density and
location , although poorly determined, are consistent with what we deduce
from the narrow emission line data. The observations also imply that the
absorber is very dense and that it covers a large fraction of the continuum
source and the BLR along the line of sight. The only additional assumption
that is required is that it also covers the line of sights to the NLR,
which is consistent with our assumption of $A=1$. The high density of the
absorber, $n > 10^{9.5}$\,cm$^{-3}$, implies that it does not contribute to
the forbidden line emission. Therefore, our gas models, which do not
include possible narrow line emission from this component, remain valid
since they rely predominantly on the forbidden lines. The analysis of Kriss
et al. (\cite{Kriss,KDZKE}) suggests that the absorber has a complex,
possibly time-variable multi-zone structure. Since the energy resolution of
the enumerated SED is too coarse, and additional free parameters will make
the enumeration too large, such detailed modeling is outside the scope of
this work. We note that the evidence for an intrinsic UV bump comes only
from the BLR covering factor problem. The absorption beyond the Lyman edge
is so high that there are almost no direct constraints on the intrinsic SED
in that energy range.

To summarize, we have to consider two possible interpretations of the
reconstructed SED of NGC\,4151.
\begin{enumerate}
\item
It is an intrinsic SED, which does not fit easily with present-day
accretion theories, which challenges AGN unification schemes, and
which photoionizes a BLR with an unusually high (perhaps even an
impossibly high) covering factor.
\item
It is an intrinsically bumpy SED (consistent with a hot thin accretion
disk), similar to the one we find in the Seyfert 2 Circinus galaxy,
which is modified by an absorber between the BLR and NLR. The
existence of an absorber with the required properties was discovered
independently by UV absorption lines.
\end{enumerate}
Both simplicity and plausibility arguments lead us to adopt the second
interpretation. While the intrinsic SED of NGC\,4151 cannot be
reconstructed from narrow emission line data because of the internal
absorption, it is encouraging that our method has succeeded in detecting
this feature in the SED. The absorption in NGC\,4151 is very complex and it
is not at all clear, due to the lack of similar quality data, how typical
is this among AGN. It is quite possible that the narrow emission lines of
other AGNs will provide us with direct information on the intrinsic
ionizing continuum. A program of ISO observations of additional AGNs,
currently in progress, will hopefully shed more light on this issue.

\section{Summary}
\label{s:summary}

In this paper we inferred the spectral shape of the obscured
photoionizing continuum source in the nucleus of the Seyfert 1 galaxy
NGC 4151.  We constrained the SED by fitting the observed intensities
of NLR emission lines, particularly the high-ionization IR lines, to a
large grid of photoionization models. The advantage of this approach
over methods that attempt to relate the observed UV and soft X-ray
spectra is that it can be used to study AGNs individually rather than
statistically, thus avoiding the problems of selection effects and
small sample statistics.

We used the available information on the line emitting gas to
construct a large grid of gas models. For each of these models, we
enumerated extensively on possible ways to bridge the UV--X-ray gap in
a SED based on simultaneous multi-band observations of the AGN
continuum, and calculated the line emission with a photoionization
code. We then employed a new method to fit the model predictions to a
compilation of observed line fluxes ranging from the UV to the IR. In
addition to minimizing over the SED, this method also minimizes over
the gas geometry and the extinction, thus making it possible to use
observational information on these quantities to further constrain the
best-fit model. Our results suggest that the filling factor of the NLR
plays an important role in determining the properties of the observed
line spectrum.  We find that the best-fit SED for NGC\,4151 {\em does
not} have a Big Blue Bump, but rather falls steeply beyond the Lyman
limit towards 4 Ryd, and then rises sharply again towards 8 Ryd. Using
our new method we confirm our previous conclusion that a Big Blue Bump
is present in the SED of the Seyfert 2 galaxy Circinus. 

We consider the possibility that the reconstructed NGC\,4151 SED is
the intrinsic SED produced by the accretion mechanism. However, such a
SED does not have enough UV photons to reproduce the BLR recombination
emission. This leads us to adopt the interpretation that the BLR is
photoionized by the intrinsic continuum source, which does contain a
strong UV component (perhaps a Big Blue Bump), but that this UV
component is absorbed by material located between the NLR and BLR. Our
analysis suggests that the absorber consists of $\sim
5\,10^{19}$\,cm$^{-2}$ of neutral hydrogen.  Such an absorber was
detected independently by UV absorption lines.

\acknowledgments

We thank Insu Yi for helpful discussions of advection dominated
accretion flows.  This work was supported by DARA under grants 50-QI-8610-8
and 50-QI-9492-3, and by the German-Israeli Foundation under grant
I-196-137.07/91.

%
%

\clearpage
\begin{deluxetable}{lr@{.}lr@{.}lr@{.}lr@{.}l}
\small
\tablecaption{The compiled emission line flux list. \label{t:flux}}
\tablewidth{0pt}
\tablehead{
\colhead{Line} &
\multicolumn{2}{c}{$\lambda_0$}&
\multicolumn{2}{c}{$E_{\rm ion}$\tablenotemark{a}}&
\multicolumn{2}{c}{$f_\ell$\tablenotemark{b}}&
\multicolumn{2}{c}{$\Delta f_\ell$\tablenotemark{c}}\\ 
\colhead{}&
\multicolumn{2}{c}{$\mu$m}&
\multicolumn{2}{c}{eV}&
\multicolumn{4}{r}{$10^{-13}{\rm erg}\,{\rm s}^{-1}{\rm cm}^{-2}$}
}
\tablecolumns{9}
\startdata
\cutinhead{UV and optical lines}
 H\,{\sc i} &     0&1216&    13&6 &     111&8  & 8&4\nl
 C\,{\sc iv}&    0&1550&    47&9 &      15&8   & 5&9\nl
 He\,{\sc ii}&   0&1640 &   54&4 &	13&6   & 2&3\nl 
 H\,{\sc i}&	 0&4861&    13&6 &       8&99  & 4&67\nl
[O\,{\sc ii}]&	 0&3726+3729&    13&6 &       31&3  & \multicolumn{2}{l}{?}\nl 
[Ne\,{\sc iii}]& 0&3868+3967&    41&0 &       23&6  & \multicolumn{2}{l}{?}\nl
[O\,{\sc iii}]&	 0&4959+5007&    35&1 &      176&0  & \multicolumn{2}{l}{?}\nl
[Fe\,{\sc vii}]& 0&5721&   100&0 &	 2&67  & \multicolumn{2}{l}{?}\nl
[Fe\,{\sc vii}]& 0&6086&   100&0 &	 4&35  & \multicolumn{2}{l}{?}\nl
 H\,{\sc i}&	 0&6563&    13&6 &        18&0 & \multicolumn{2}{l}{?}\nl
[Ar\,{\sc iii}]& 0&7136&    27&6 &	 2&96  & \multicolumn{2}{l}{?}\nl
[O\,{\sc ii}]&   0&7320+7330&	 13&6 &	 3&85  & \multicolumn{2}{l}{?}\nl
[S\,{\sc iii}]&	 0&9069+9531&	 23&3 &	 27&3  & 3&4\nl  
\cutinhead{ISO IR lines}
[Mg\,{\sc viii}]& 3&028&     224&9 &	 0&62  & 0&13\nl
[Si\,{\sc ix}]&	  3&936&     303&2 &	 0&41  & 0&08\nl
[Ne\,{\sc vi}]&	  7&642&     126&2 &	 7&9   & 1&6\nl    
[Ar\,{\sc iii}]&  8&991&      27&6 &	 2&20  & 0&46\nl
[S\,{\sc iv}]&   10&51&       34&8 &	 11&3  & 2&3\nl    
[Ne\,{\sc ii}]&  12&81&       21&6 &	 11&8  & 2&4\nl    
[Ne\,{\sc v}]&   14&32&       97&1 &      5&5  & 1&2\nl    
[Ne\,{\sc iii}]& 15&56&       41&0 &	 20&7  & 4&1\nl    
[S\,{\sc iii}]&  18&71&       23&3 &      5&4  & 1&1\nl    
[Ne\,{\sc v}]&   24&32&       97&1 &      5&6  & 1&1\nl    
[O\,{\sc iv}]&   25&89&       54&9 &     20&3  & 4&1\nl    
[S\,{\sc iii}]&  33&48&       23&3 &	  8&1  & 1&7\nl    
[Ne\,{\sc iii}]& 36&01&       41&0 &	  3&5  & 0&7\nl
[O\,{\sc iii}]&  51&81&       35&1 &	 10&3  & 3&2\nl    
[O\,{\sc iii}]&  88&36&       35&1 &	  6&8  & 0&6\tablebreak
\cutinhead{Problematic lines not used in fit}
 He\,{\sc ii}&   0&4686 &    54&4 &	 2&05  & 0&8\nl
[Ar\,{\sc iv}]&  0&4711 &    40&7 &	 1&25  & \multicolumn{2}{l}{?}\nl
[N\,{\sc ii}]&   0&6548+6584&  14&5 &	18&6   & \multicolumn{2}{l}{?}\nl
[S\,{\sc ii}]&   0&6717 &    10&4 &	14&6   & \multicolumn{2}{l}{?}\nl
[S\,{\sc ii}]&   0&6731 &    10&4 &	17&4   & \multicolumn{2}{l}{?}\nl
[Ar\,{\sc v}]&   0&7006 &    59&8 &	 0&46  & 0&08\nl
[Ar\,{\sc iii}]& 0&7751 &    27&6 &	 0&89  & \multicolumn{2}{l}{?}\nl
[Si\,{\sc vii}]& 2&481  &   205&1 &	 1&2   & 0&1\nl
[Si\,{\sc ix}]&  2&585  &    303&2 &	 0&23  & 0&07\nl
[Mg\,{\sc iv}]&  4&49   &     80&1 &	 0&31  & 0&07\nl
[Mg\,{\sc vii}]& 5&50   &   186&5 &     $<$1&0 & \multicolumn{2}{l}{---}\nl
[Mg\,{\sc v}]&   5&608  &   109&2 &     $<$1&5 & \multicolumn{2}{l}{---}\nl
[Si\,{\sc ii}]& 34&81   &     8&2 &	 15&6  & 3&3\nl    
[O\,{\sc i}]&   63&18   &     0&0 &	 30&6  & 3&1\nl    
[O\,{\sc i}]&  145&5    &     0&0 &	 1&94  & 0&05\nl
[C\,{\sc ii}]& 157&7    &    11&3 &	 5&69  & 0&35\nl
\enddata
\tablenotetext{a}{The ionization energy required to produce the
emitting ion from the preceding ionization stage.}
\tablenotetext{b}{Observed flux. For permitted lines, the flux of the
decomposed narrow component is quoted.}
\tablenotetext{c}{Error estimate on observed flux. The ISO-SWS errors
were estimated by adding in quadrature the error due to the uncertainty
in defining the underlying continuum and a global 20\% statistical
error. A question mark means that error estimates are unavailable.}
\end{deluxetable} 

\clearpage
\begin{deluxetable}{llllllllll}
\tablecaption{The best-fit scores for 
the single component gas models. The best-fit model is emphasized in bold
font. \label{t:score}}
\small
\tablewidth{0pt}
\tablehead{
\multicolumn{3}{r}{$U$}   & 
\multicolumn{1}{c}{0.005} &
\multicolumn{1}{c}{0.01}  &
\multicolumn{1}{c}{0.025} &
\multicolumn{1}{c}{0.05}  &
\multicolumn{1}{c}{0.1}   &
\multicolumn{1}{c}{0.2}   &
\multicolumn{1}{c}{0.4}   \\
$\log r_0$& $F$   &\multicolumn{8}{c}{}
}
\tablecolumns{10}
\startdata
\cutinhead{ISO fit}
  17.5   &1     & $S$
& {\em 2.32}  & {\em 1.75}  & {\em 1.64}  & {\em 1.60}  & {\em 1.59}  & {\em 1.66}  & {\em 1.73}  \nl
         &        &$\maxs$\tablenotemark{a}
& 6.32$^{-1}$ & 2.48$^{-1}$ & 2.35$^{-1}$ & 2.33$^{-1}$ & 2.09$^{-1}$ & 2.56        & 3.47\nl
         &        & $C$    
& 0.21        & 0.23        & 0.31        & 0.40        & 0.39        & 0.42        & 0.51\nl
         &        &$\theta$\tablenotemark{b}
& 1.54        & 1.02        & 0.66        & 0.46        & 0.36        & 0.27        & 0.18\nl
         &        &$\tion$\tablenotemark{c}
& 0.01        & 0.01        & 0.01        & 0.01        & 0.01        & 0.02        & 0.03\nl
         &        & $\EBV$
& 0.00        & 0.00        & 0.00        & 0.00        & 1.59        & 3.42        & 4.55\nl
  $\infty$   &1     & $S$
& {\em 2.06}  & {\em 1.80}  & {\em 1.78}  & {\em 1.76}  & {\em 1.92}  & {\em 2.29}  & {\em 2.25}  \nl
         &        &$\maxs$
& 3.29$^{-1}$ & 2.53$^{-1}$ & 2.55$^{-1}$ & 2.56$^{-1}$ & 3.35        & 5.45        & 7.20\nl
         &        & $C$    
& 0.17        & 0.16        & 0.20        & 0.29        & 0.28        & 0.28        & 0.33\nl
         &        &$\theta$
& 1.54        & 1.15        & 0.73        & 0.48        & 0.34        & 0.26        & 0.15\nl
         &        &$\tion$
& 0.06        & 0.12        & 0.17        & 0.20        & 0.24        & 0.25        & 0.41\nl
         &        & $\EBV$
& 0.00        & 0.00        & 0.00        & 0.00        & 0.09        & 2.26        & 3.25\nl
  $\infty$   &0.065     & $S$
& {\em 2.10}  & {\em 1.81}  & {\em 1.82}  & {\em 1.80}  & {\em 1.93}  & {\em 2.06}  & {\em 2.10}  \nl
         &        &$\maxs$
& 3.32$^{-1}$ & 2.48$^{-1}$ & 3.07$^{-1}$ & 2.71$^{-1}$ & 3.55$^{-1}$ & 4.99        & 5.64\nl
         &        & $C$
& 0.18        & 0.22        & 0.39        & 0.36        & 0.33        & 0.38        & 0.40\nl
         &        &$\theta$
& 1.54        & 1.08        & 0.60        & 0.45        & 0.34        & 0.23        & 0.16\nl
         &        &$\tion$
& 0.56        & 0.72        & 0.65        & 0.74        & 0.78        & 0.86        & 0.87\nl
         &        & $\EBV$
& 0.00        & 0.00        & 0.00        & 0.00        & 0.00        & 1.14        & 1.57\nl
  $\infty$   &0.00065     & $S$
& {\em 2.87}  & {\em 2.06}  & {\bf 1.80}  & {\em 1.79}  & {\em 1.79}  & {\em 1.79}  & {\em 1.79}  \nl
         &        &$\maxs$
& 13.49$^{-1}$ & 3.19$^{-1}$ &{\bf 2.93$^{-1}$}& 3.07$^{-1}$ & 3.11$^{-1}$ & 3.12$^{-1}$ & 3.11$^{-1}$ \nl
         &        & $C$
& 0.22        & 0.28        &{\bf 0.29}   & 0.28        & 0.28        & 0.26        & 0.28\nl
         &        &$\theta$
& 1.62        & 1.05        &{\bf 0.64}   & 0.46        & 0.32        & 0.24        & 0.16\nl
         &        &$\tion$
& 4.38        & 4.51        &{\bf 4.68}   & 4.85        & 4.95        & 4.95        & 5.01\nl
         &        & $\EBV$
& 0.00        & 0.00        &{\bf 0.00}   & 0.00        & 0.00        & 0.00        & 0.00\nl
\tablebreak
\cutinhead{Full fit}
  17.5   &1     & $S$
& {\em 2.02}  & {\em 1.80}  & {\em 1.87}  & {\em 1.96}  & {\em 2.10}  & {\em 2.33}  & {\em 2.36}  \nl
         &        &$\maxs$
& 5.68$^{-1}$ & 2.86        & 3.38$^{-1}$ & 3.81$^{-1}$ & 4.65$^{-1}$ & 5.04        & 5.92\nl
         &        & $C$
& 0.23        & 0.26        & 0.38        & 0.47        & 0.48        & 0.47        & 0.45\nl
         &        &$\theta$
& 1.54        & 1.02        & 0.63        & 0.46        & 0.32        & 0.23        & 0.16\nl
         &        &$\tion$
& 0.01        & 0.01        & 0.01        & 0.01        & 0.02        & 0.02        & 0.03\nl
         &        & $\EBV$
& 0.07        & 0.04        & 0.03        & 0.02        & 0.01        & 0.01        & 0.00\nl
  $\infty$   &1     & $S$
& {\em 1.95}  & {\em 1.90}  & {\em 2.05}  & {\em 2.15}  & {\em 2.35}  & {\em 2.82}  & {\em 2.69}  \nl
         &        &$\maxs$
& 3.28$^{-1}$ & 2.81$^{-1}$ & 3.46$^{-1}$ & 3.80$^{-1}$ & 5.81$^{-1}$ & 13.87        & 13.00\nl
         &        & $C$
& 0.23        & 0.24        & 0.31        & 0.45        & 0.38        & 0.22        & 0.27\nl
         &        &$\theta$
& 1.48        & 1.04        & 0.65        & 0.43        & 0.32        & 0.21        & 0.15\nl
         &        &$\tion$
& 0.07        & 0.14        & 0.19        & 0.18        & 0.24        & 0.47        & 0.52\nl
         &        & $\EBV$
& 0.05        & 0.04        & 0.02        & 0.01        & 0.00        & 0.00        & 0.00\nl
  $\infty$   &0.065     & $S$
& {\em 1.94}  & {\em 1.86}  & {\em 2.03}  & {\em 2.15}  & {\em 2.35}  & {\em 2.55}  & {\em 2.53}  \nl
         &        &$\maxs$
& 3.30$^{-1}$ & 2.64$^{-1}$ & 3.05$^{-1}$ & 3.79$^{-1}$ & 5.45$^{-1}$ & 11.09        & 11.04\nl
         &        & $C$
& 0.25        & 0.26        & 0.41        & 0.52        & 0.46        & 0.37        & 0.38\nl
         &        &$\theta$
& 1.48        & 1.04        & 0.62        & 0.43        & 0.32        & 0.22        & 0.15\nl
         &        &$\tion$
& 0.61        & 0.79        & 0.87        & 0.76        & 0.83        & 1.06        & 1.12\nl
         &        & $\EBV$
& 0.05        & 0.04        & 0.03        & 0.01        & 0.00        & 0.00        & 0.00\nl
  $\infty$   &0.00065     & $S$
& {\em 2.31}  & {\em 1.95}  & {\bf 1.90}  & {\em 1.91}  & {\em 1.92}  & {\em 1.92}  & {\em 1.93}  \nl
         &        &$\maxs$
& 12.45$^{-1}$ & 2.98        &{\bf 3.03$^{-1}$} & 3.49$^{-1}$ & 3.53$^{-1}$ & 3.16$^{-1}$ & 3.16$^{-1}$ \nl
         &        & $C$
& 0.24        & 0.26        &{\bf 0.24}  & 0.28        & 0.28        & 0.23        & 0.23\nl
         &        &$\theta$
& 1.62        & 1.09        &{\bf 0.69}  & 0.47        & 0.33        & 0.24        & 0.17\nl
         &        &$\tion$
& 4.38        & 4.61        &{\bf 4.95}  & 4.91        & 5.01        & 5.25        & 5.31\nl
         &        & $\EBV$
& 0.08        & 0.05        &{\bf 0.03}  & 0.02        & 0.02        & 0.03        & 0.03\nl
\enddata
\tablenotetext{a}{When the model to data ratio is smaller than 1, the
discrepancy is given as the reciprocal of the ratio.}
\tablenotetext{b}{Angular distance of cloud surface from center, in
arcsec, assuming $A=1$.}
\tablenotetext{c}{Angular thickness of the ionized layer in the cloud, in
arcsec, assuming $h=0.65$.}
\end{deluxetable}

\clearpage
\begin{deluxetable}{lll}
\small
\tablecaption{The best-fit results of the SED and cloud models. \label{t:fit}}
\tablewidth{0pt}
\tablehead{
\colhead{} & \colhead{ISO fit (15 lines)} & \colhead{Full fit (28 lines) }
}
\tablecolumns{3}
\startdata
\cutinhead{Input parameters}
$U$          &  0.025                &  0.025               \nl
$n_0$\tablenotemark{a}&
                $10^3$               &  $10^3$              \nl
$\log r_0$   &  $\infty$             &  $\infty$            \nl
$A$          &  $1$                  &  $1$                 \nl
$F$          &  $6.5\,10^{-4}$       &  $6.5\,10^{-4}$      \nl
$h$          &  $0.65$               &  $0.65$              \nl
$\log L_1$\tablenotemark{b}& 
                27.88                &  27.88               \nl
$\log L_{4134}$&
                24.18                &  24.18               \nl
\cutinhead{Best-fit results\tablenotemark{c}}
$S$          &  1.8                  &  1.9                 \nl
$\max S_\ell$\tablenotemark{d}&  
                2.93$^{-1}$          &  3.03$^{-1}$         \nl
worst line   &$\bNeIIIbB$            &  [Fe\,{\sc vii}]\,$\lambda 6086$\AA\nl
$\log L_4$   &  25.4 (25.4, 26.2)    &  25.4 (25.4, 26.6)   \nl
$\log L_8$   &  26.6 (26.2, 27.0)    &  27.0 (26.4, 27.6)   \nl
$\log L_{30}$&  26.0 (25.7, 26.0)    &  26.0 (25.7, 26.0)   \nl
$C$          &  0.29 (0.17, 0.43)    &  0.24 (0.10, 0.40)   \nl
$\theta$\tablenotemark{e}&
                0.64 (0.62, 0.78)    &  0.69 (0.63, 0.89)   \nl
$\theta_{\rm ion}$\tablenotemark{f}&
                4.68 (3.84, 4.95)    &  4.95 (4.14, 5.35)   \nl
$\EBV$       &  0.00 (0.00, 2.41)    &  0.03 (0.00, 0.20)   \nl
$\qion$\tablenotemark{g}&
                7.35                 &  8.36                \nl
$\log Q_{\rm ion}$\tablenotemark{h}&
               53.31                 &  53.37               \nl
$\langle h\nu \rangle$\tablenotemark{i}&
                3.2                  &   4.1                \tablebreak
\cutinhead{Residual correlations\tablenotemark{j}}\nl
$\lambda_0$  &$\tau=+0.24$ $P_0=0.22$ &$\tau=+0.11$ $P_0=0.43$\nl
$\Eion$      &$\tau=-0.11$ $P_0=0.58$ &$\tau=-0.02$ $P_0=0.87$\nl
depletion    &$\tau=+0.00$ $P_0=1.00$ &$\tau=+0.03$ $P_0=0.91$\nl
\enddata
\tablenotetext{a}{Hydrogen density in cm$^{-3}$.}
\tablenotetext{b}{Luminosity in erg\,s$^{-1}$\,Hz$^{-1}$, 
using $L_\nu=4\pi D^2 f_\nu$, with $D=20$\,Mpc.}
\tablenotetext{c}{Values in parentheses are the 99.9\% confidence intervals.}
\tablenotetext{d}{When the model to data ratio is smaller than 1, the
discrepancy is given as the reciprocal of the ratio.}
\tablenotetext{e}{Angular distance of cloud surface from center, in
arcsec, assuming $A=1$. ($1\arcsec=73.8$\,pc for $h=0.65$)}
\tablenotetext{f}{Angular thickness of the ionized layer in the cloud, in
arcsec, assuming $h=0.65$.}
\tablenotetext{g}{Ionizing photon flux in s$^{-1}$\,cm$^{-2}$.}
\tablenotetext{h}{Ionizing photon luminmosity in s$^{-1}$,
assuming isotopic emission and $h=0.65$.}
\tablenotetext{i}{Mean ionizing photon energy in Ryd, calculated between 1
and 30 Ryd.}
\tablenotetext{j}{$P_0$ is the probability for a correlation $\tau$ in 
random data.}
\end{deluxetable}

%
%

\setcounter{figure}{0}
\clearpage
\begin{figure}
\plotone{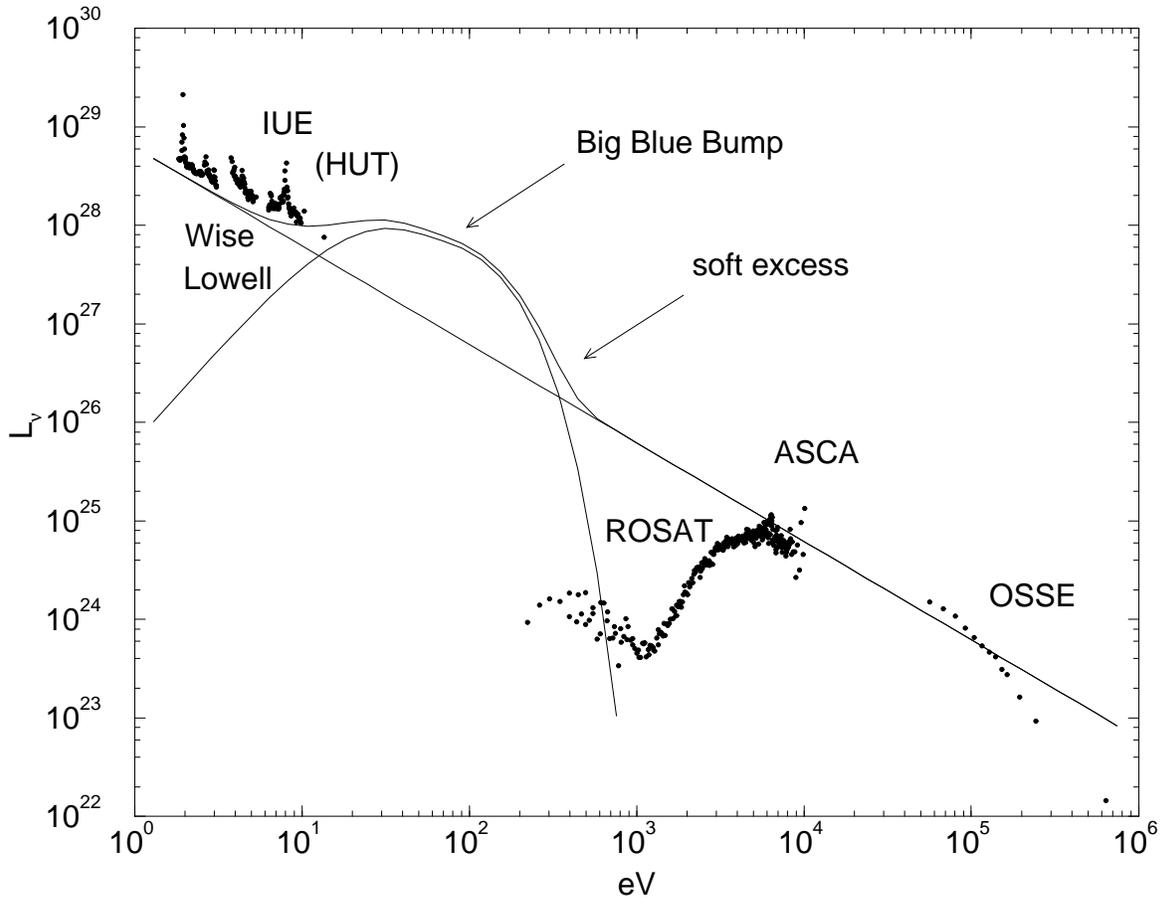}
\caption{A nearly simultaneous SED of NGC\,4151 from the optical to
$\gamma$-rays (points) (From Edelson et al. (\cite{Edelson}); HUT result
from Kriss et al. \cite{KDZKE}). The observed fluxes were translated to
luminosity assuming isotropic emission and a distance of $20$\,Mpc. A sketch
of a bump superimposed on a power-law is presented for illustration
purposes (line).
\label{f:sed+bump}
}
\end{figure}

\clearpage
\begin{figure}
\plotone{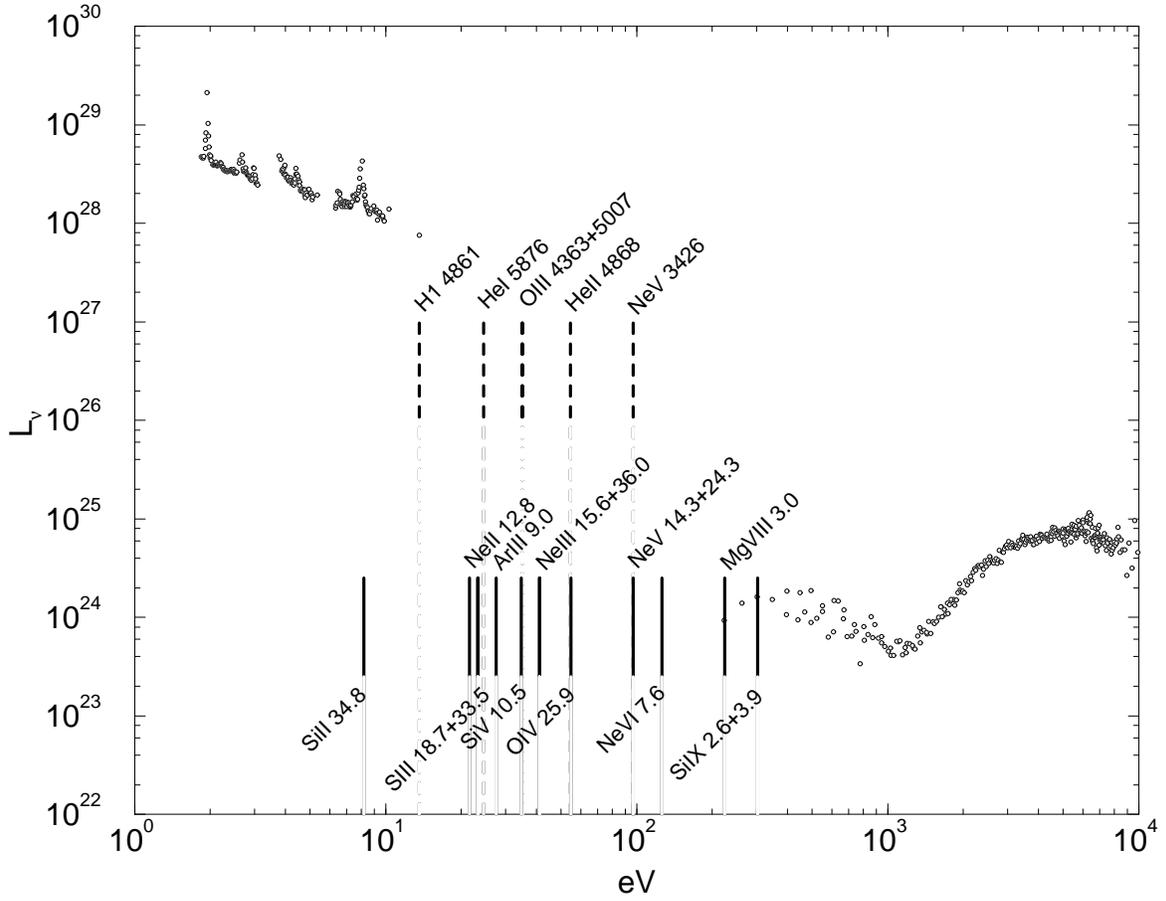}
\caption{The strong observed IR emission lines (full lines) and some of the
optical emission lines (dashed lines) plotted on the energy axis
according the ionization energy required to produce the emitting ion
from the preceding ionization stage. Also shown is the observed SED of
NGC\,4151 (points).
\label{f:eion}
}
\end{figure}

\clearpage
\begin{figure} 
\epsscale{.75}
\plotone{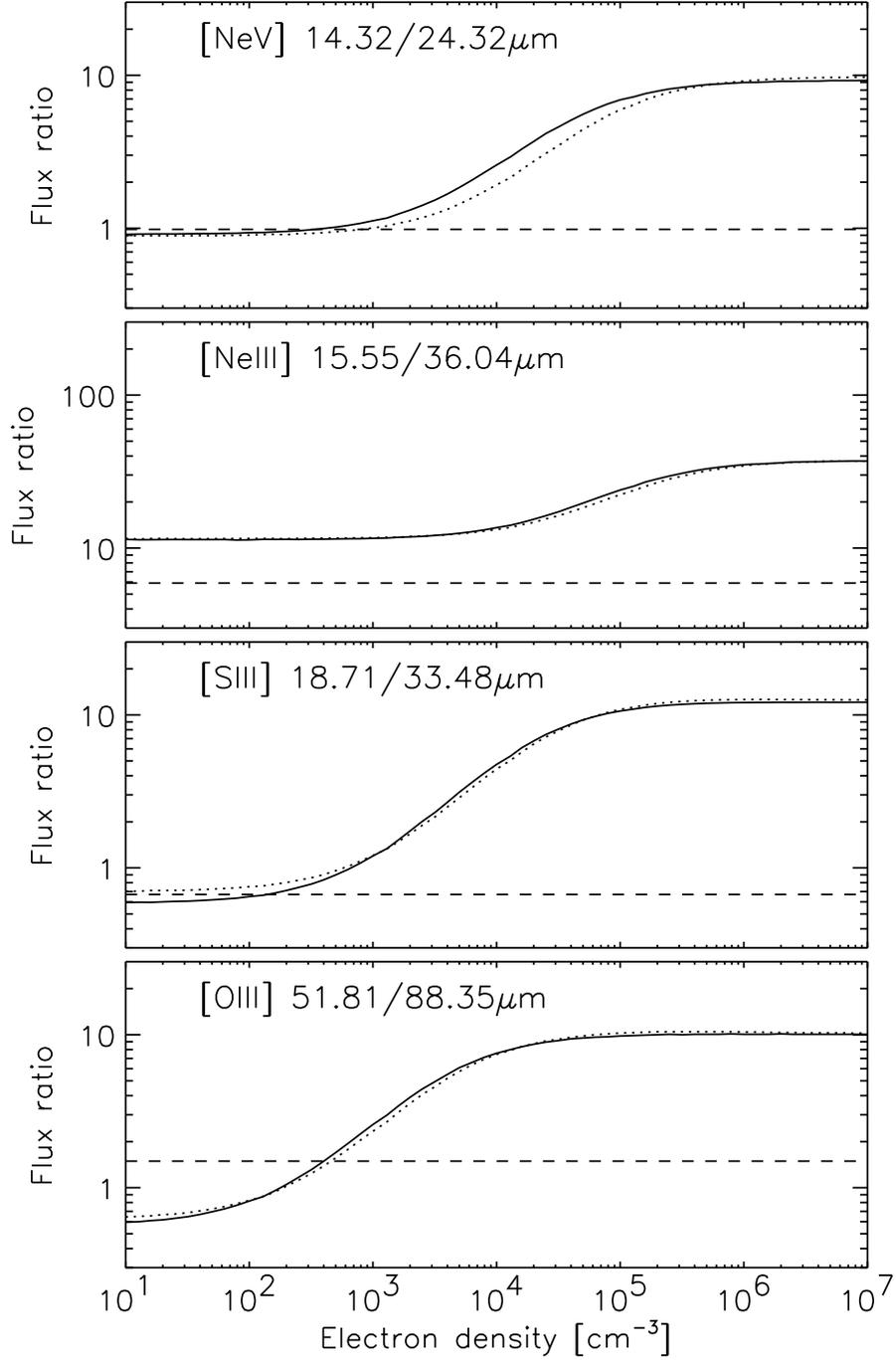} 	
\caption{Density-sensitive ratios of fine structure transitions from the
ground state triplets of [Ne\,{\sc v}], [Ne\,{\sc iii}], [S\,{\sc iii}],
and [O\,{\sc iii}]. Full lines refer to an electron temperature of
10000\,K, dotted lines to 20000\,K. The dashed lines indicate the 
observed ratios. 
\label{f:density}
}
\end{figure} 

\clearpage
\begin{figure} 
\plotone{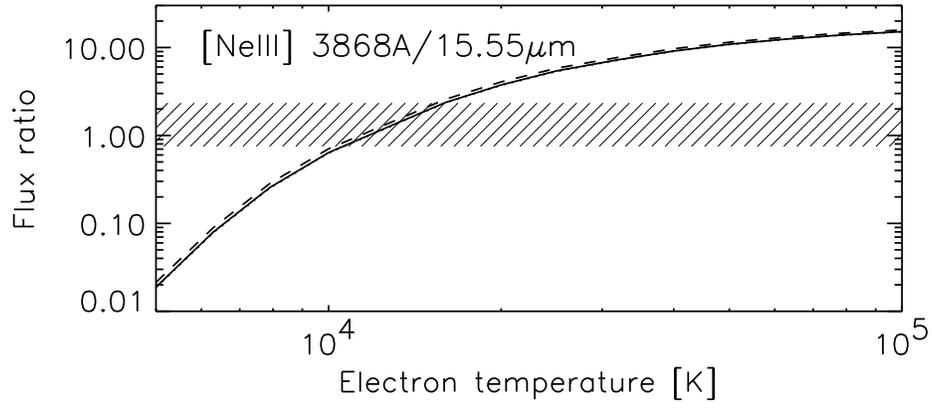} 	
\caption{Temperature-sensitive optical/IR line ratio for lines of
[Ne\,{\sc iii}], computed for electron densities of 100 and 1000 
(full line) and 10000\,cm$^{-3}$ (dashed line). The hashed area indicates
the uncertainty range of the extinction corrected observed ratio.
\label{f:temp}
}
\end{figure} 

\clearpage
\begin{figure}
\plotone{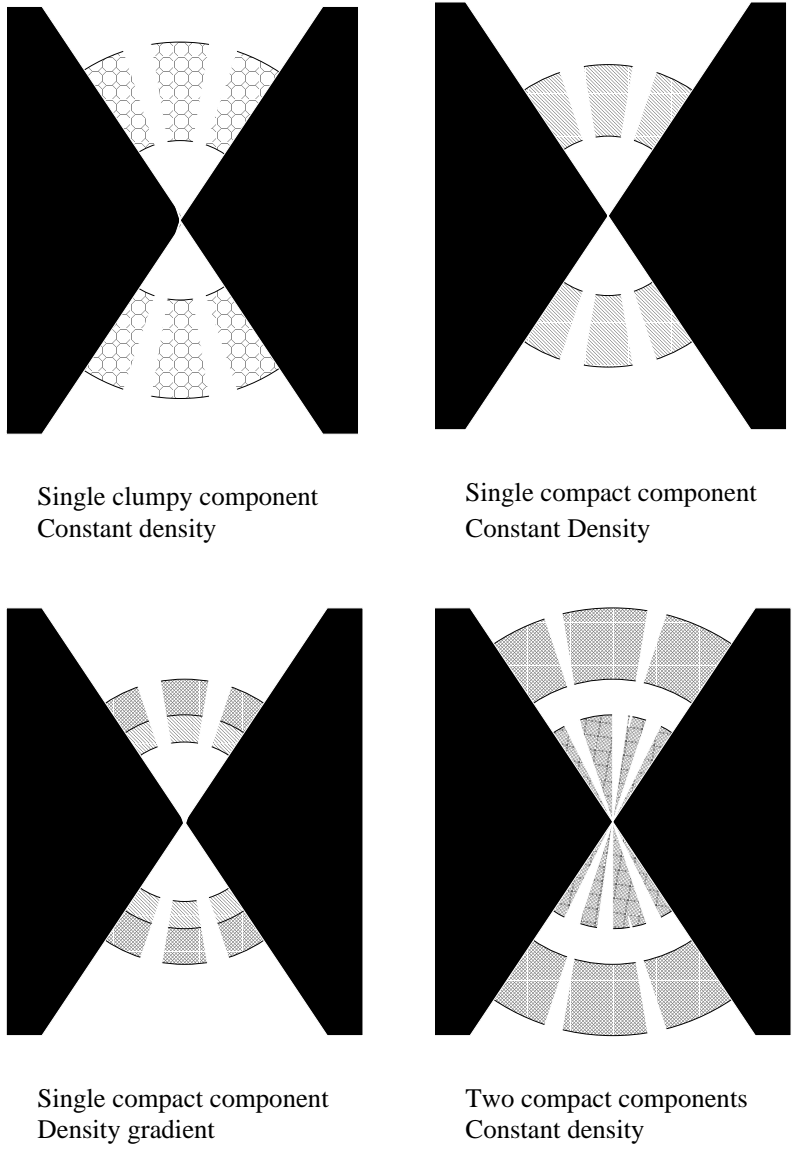} 
\caption{The four gas configurations considered in this work. The black
region corresponds to the regions outside the illuminated cone. The
continuum source is at the center.
\label{f:geom}
}
\end{figure} 

\clearpage
\begin{figure}
\plotone{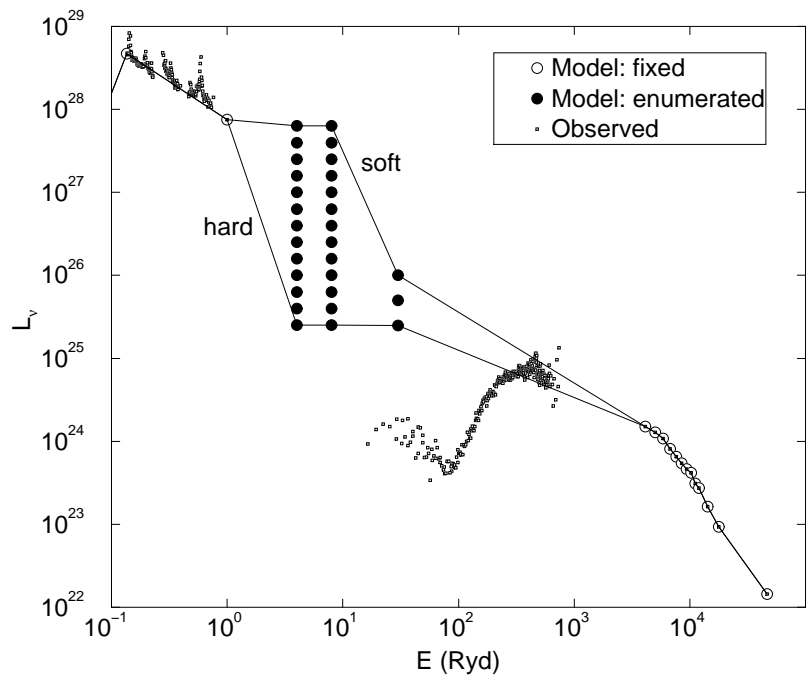} 
\caption{The template used for enumerating on the SED
of NGC\,4151. The most and least luminous SEDs are indicated by the
lines, together with the relative sense of their spectral hardness
between 1 and 30 Ryd. 
\label{f:sed_temp}
}
\end{figure} 

\clearpage
\begin{figure}
\plotone{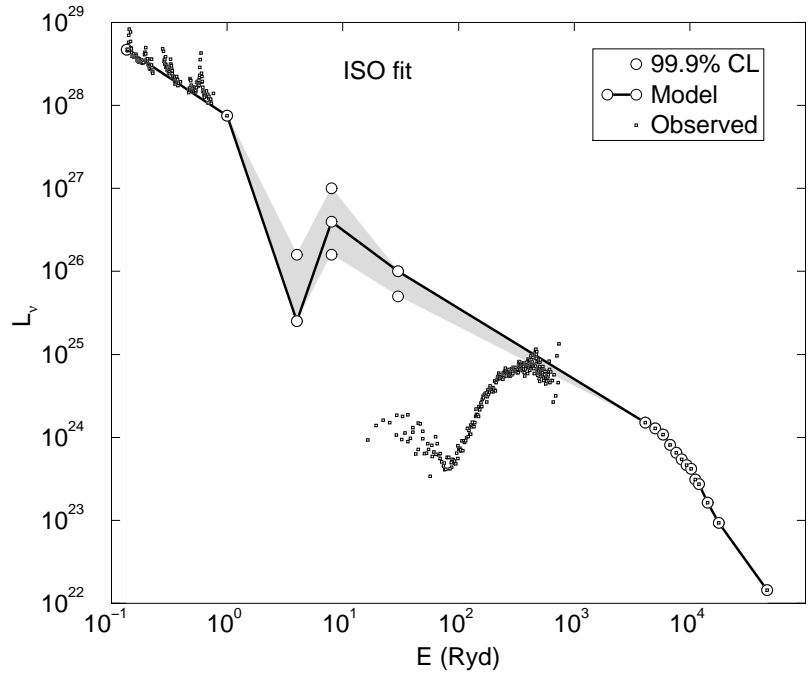}
\caption{The best-fit SED models for the ISO fit and full fit (line). 
The shaded area is the 99.9 \% confidence limit on the SED. 
Both models have $U=0.025$, $F=6.5\,10^{-4}$ and constant density.
\label{f:best_SED}
}
\end{figure}  
\clearpage
\addtocounter{figure}{-1}
\begin{figure}
\plotone{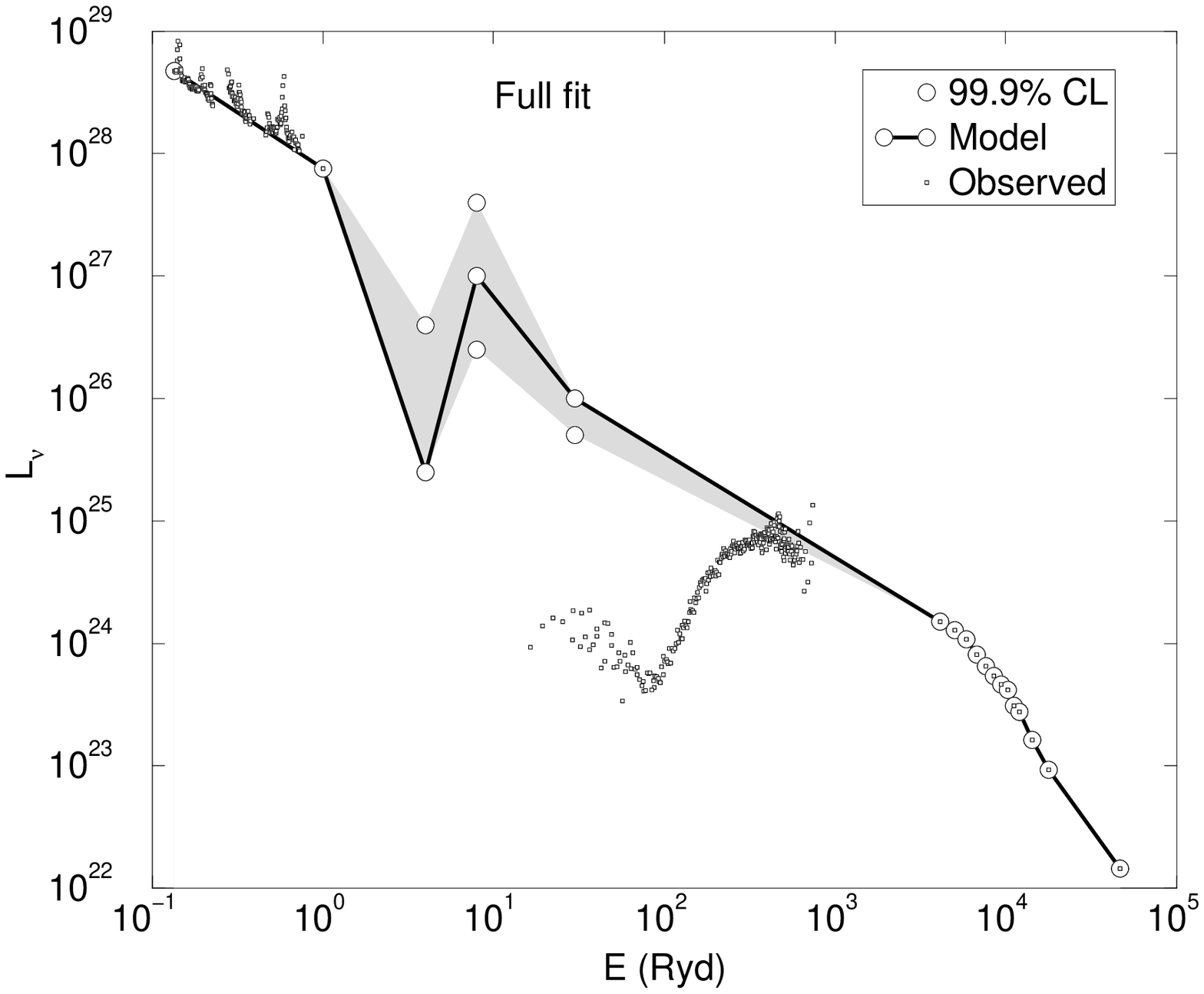}
\caption{Continued.}
\end{figure}  

\clearpage
\begin{figure}
\plotone{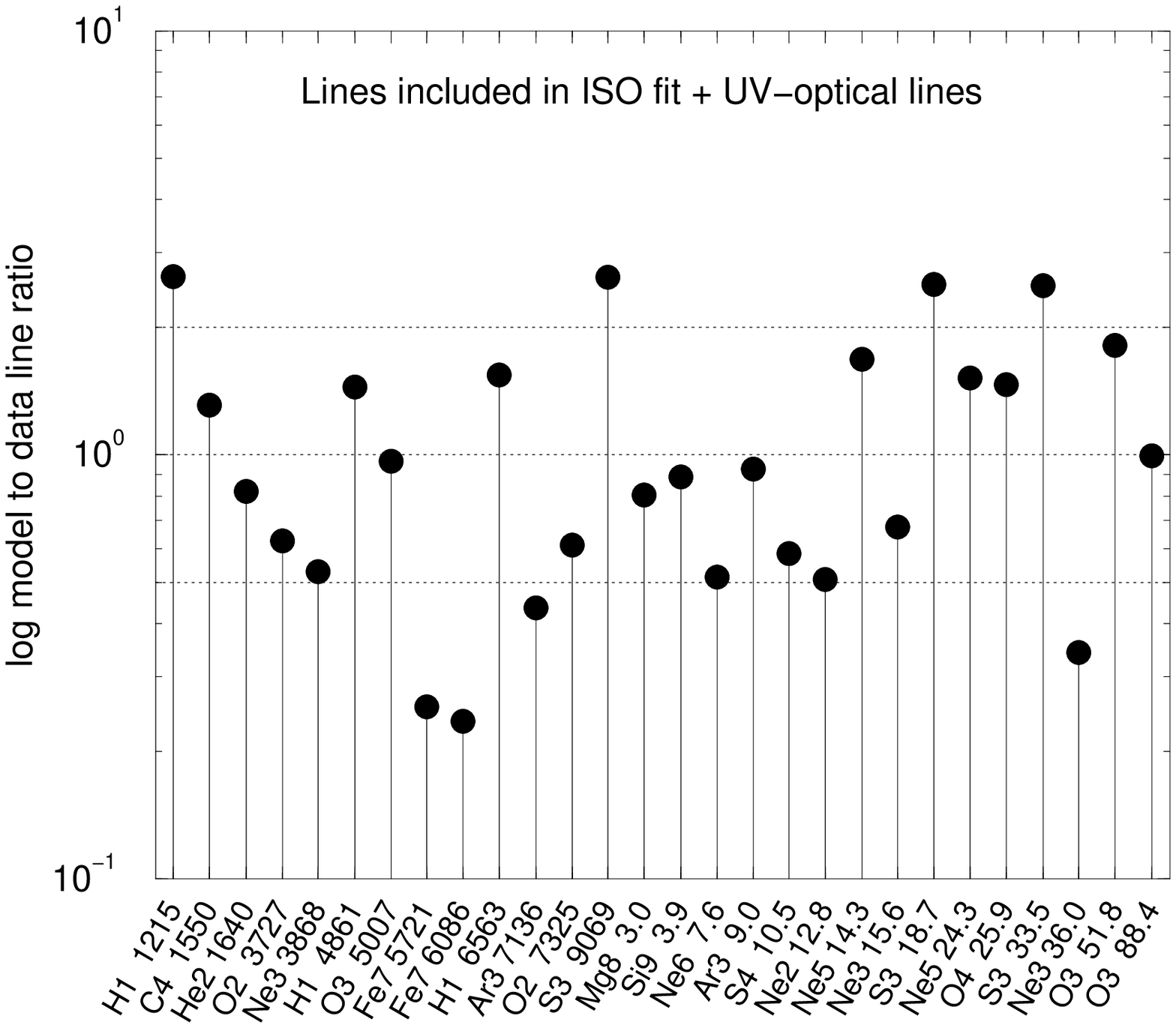}
\caption{The model to data line ratios in the best-fit model based on the
ISO lines only ($U=0.025$, $F=6.5\,10^{-4}$, constant density).
\label{f:ratio_ISO}
}
\end{figure}
\clearpage
\addtocounter{figure}{-1}
\begin{figure}
\plotone{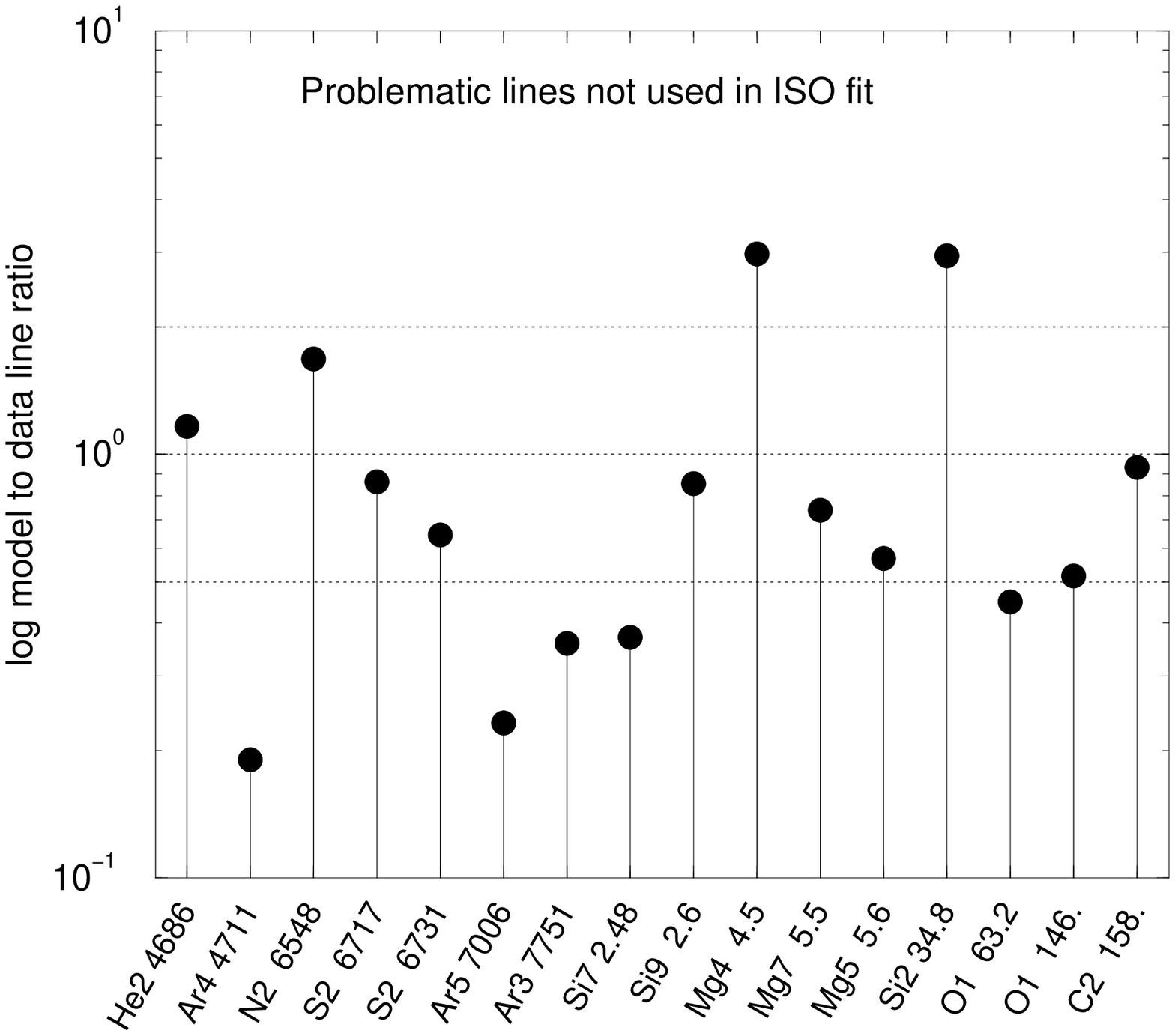}
\caption{Continued.}
\end{figure}  

\clearpage
\begin{figure}
\plotone{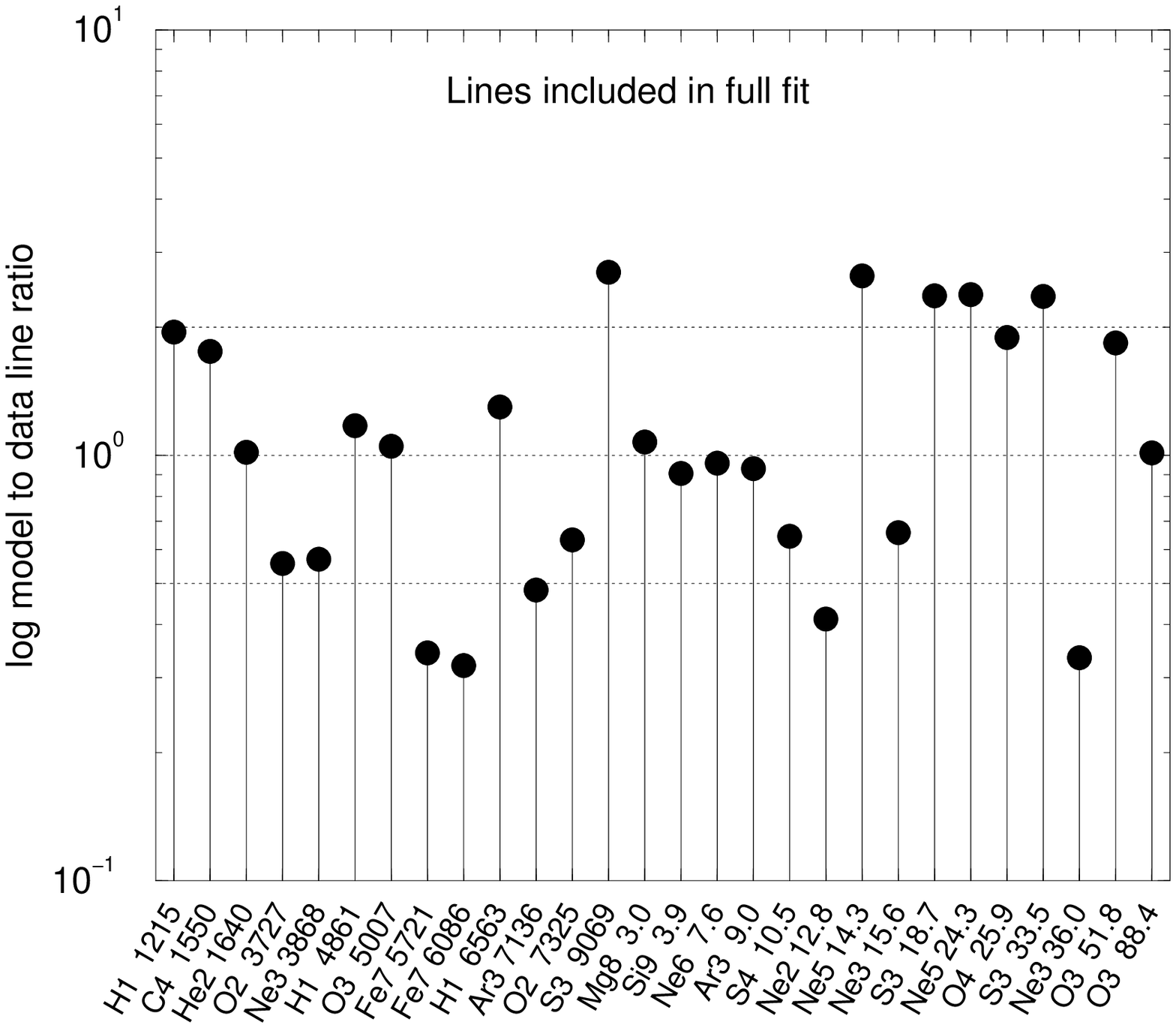}
\caption{The model to data line ratios in the best-fit model
based on the full line list ($U=0.025$, $F=6.5\,10^{-4}$,
constant density).
\label{f:ratio_full}
} 
\end{figure}  
\clearpage
\addtocounter{figure}{-1}
\begin{figure}
\plotone{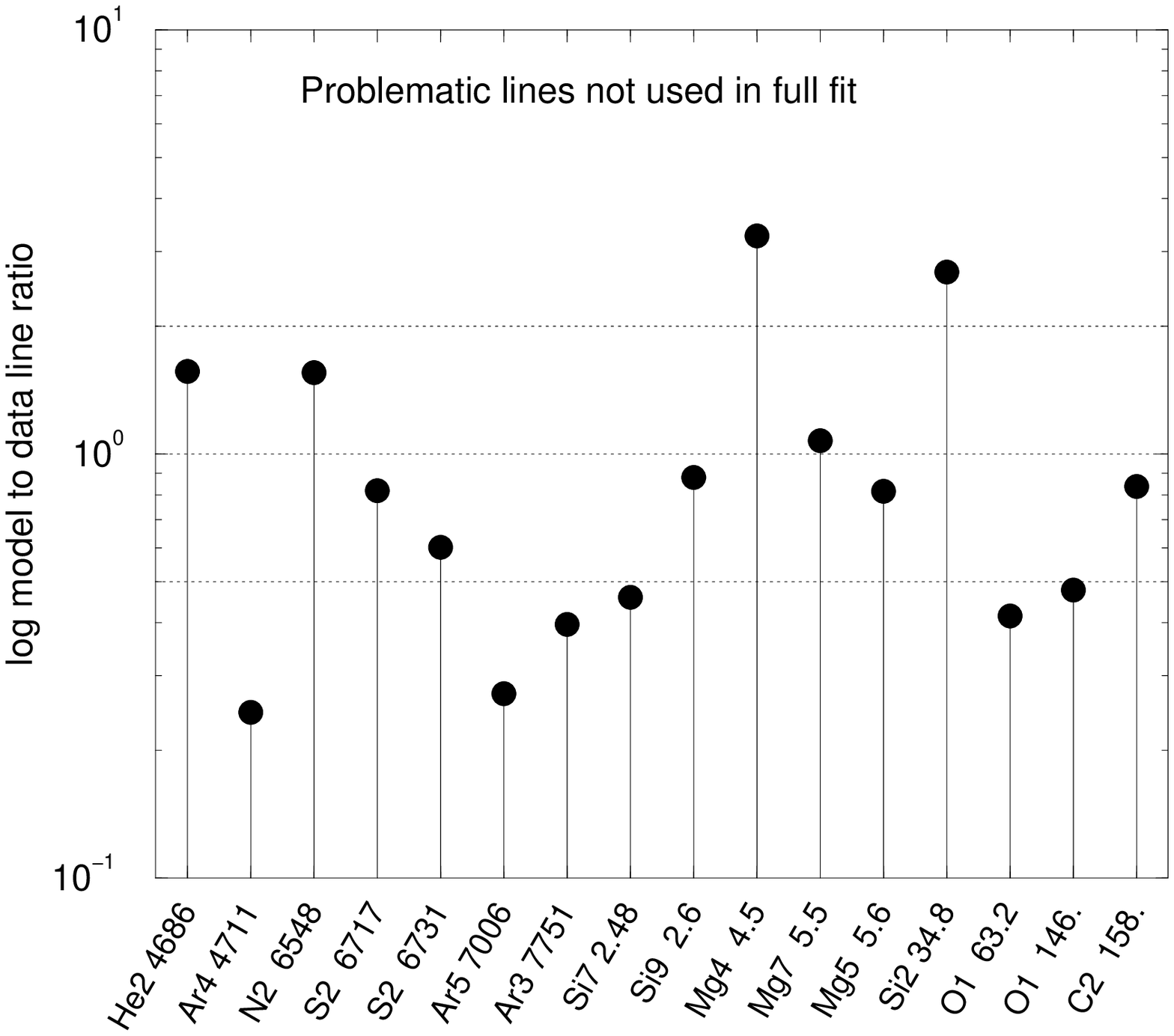} 
\caption{Continued.}
\end{figure}  

\clearpage
\begin{figure}
   \begin{tabular}{c}
   \rotate[r]{\epsfxsize=250pt
              \epsfbox{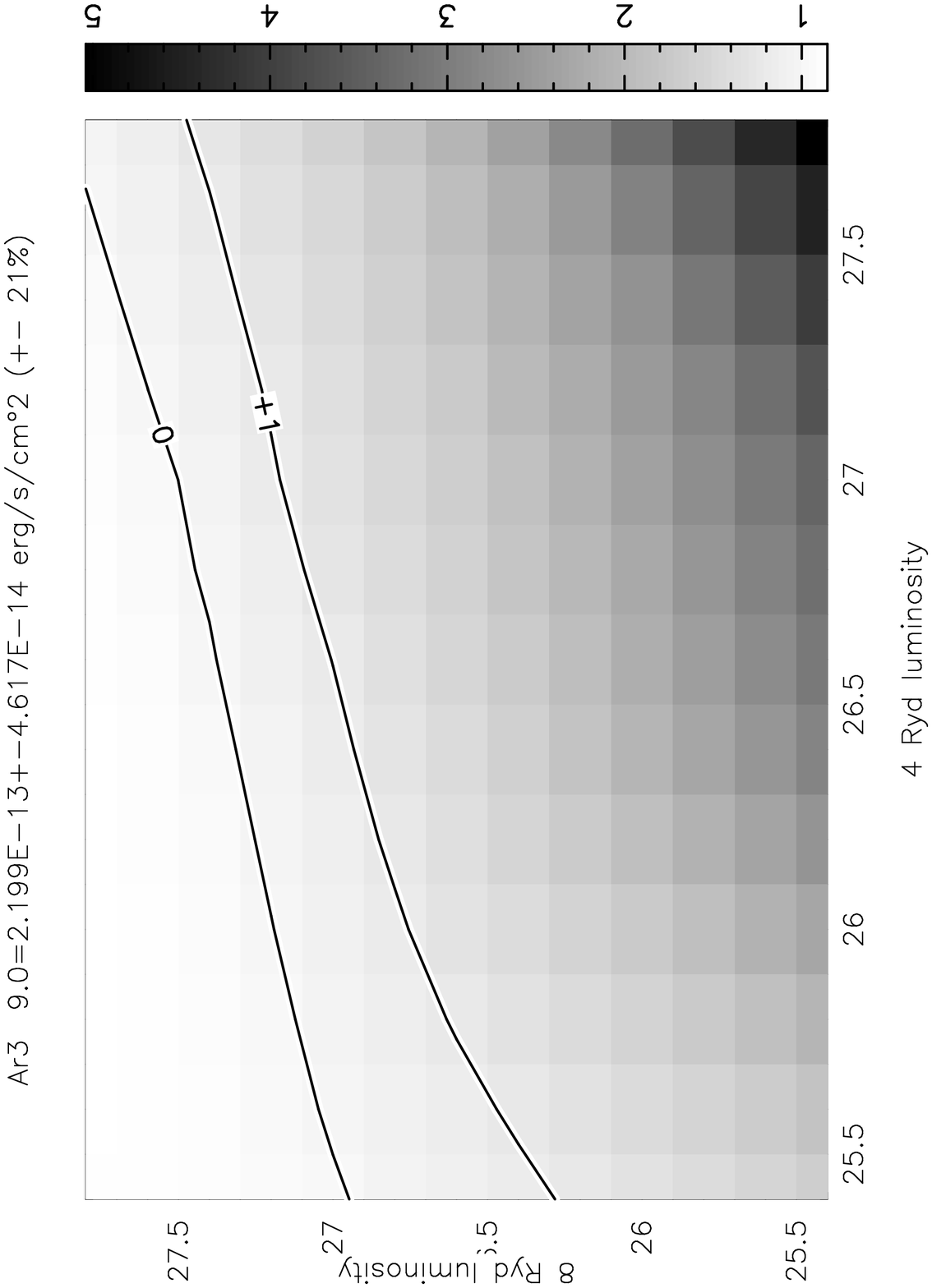}
             }\\
   \end{tabular}
\caption{The model to data ratio of $\bArIIIb$ and $\bSiIXbB$ as function
of $L_4$ and $L_8$ (for fixed $\log L_{30} = 26.0$) in the best full fit
model ($U=0.025$, $F=6.5\,10^{-4}$, constant density). The contours marked
by 0, +1 and -1 represent the observed value and the upper and lower errors
on it, respectively (when either of the $\pm1$ contours isn't shown, it
lies outside the enumerated region). The lower right hand side corner
corresponds to the case of an extreme Big Blue Bump that peaks at around 4
Ryd and then falls steeply towards 8 Ryd.
\label{f:gray}
}
\end{figure}  
\clearpage
\addtocounter{figure}{-1}
\begin{figure}
   \begin{tabular}{c}
   \rotate[r]{\epsfxsize=250pt
              \epsfbox{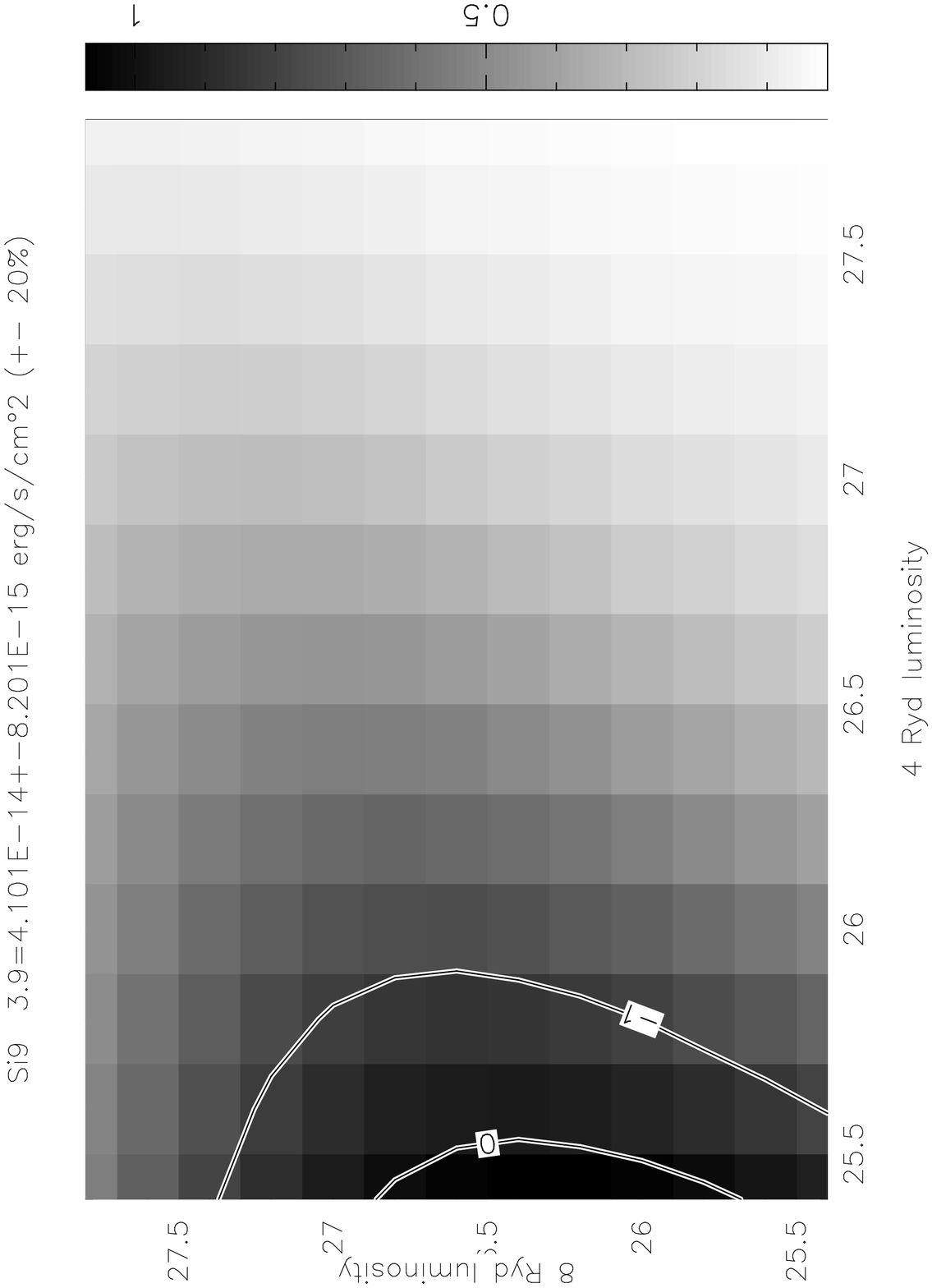} 	
             }
   \end{tabular}
\caption{Continued.}
\end{figure}  

\clearpage
\begin{figure}
\plotone{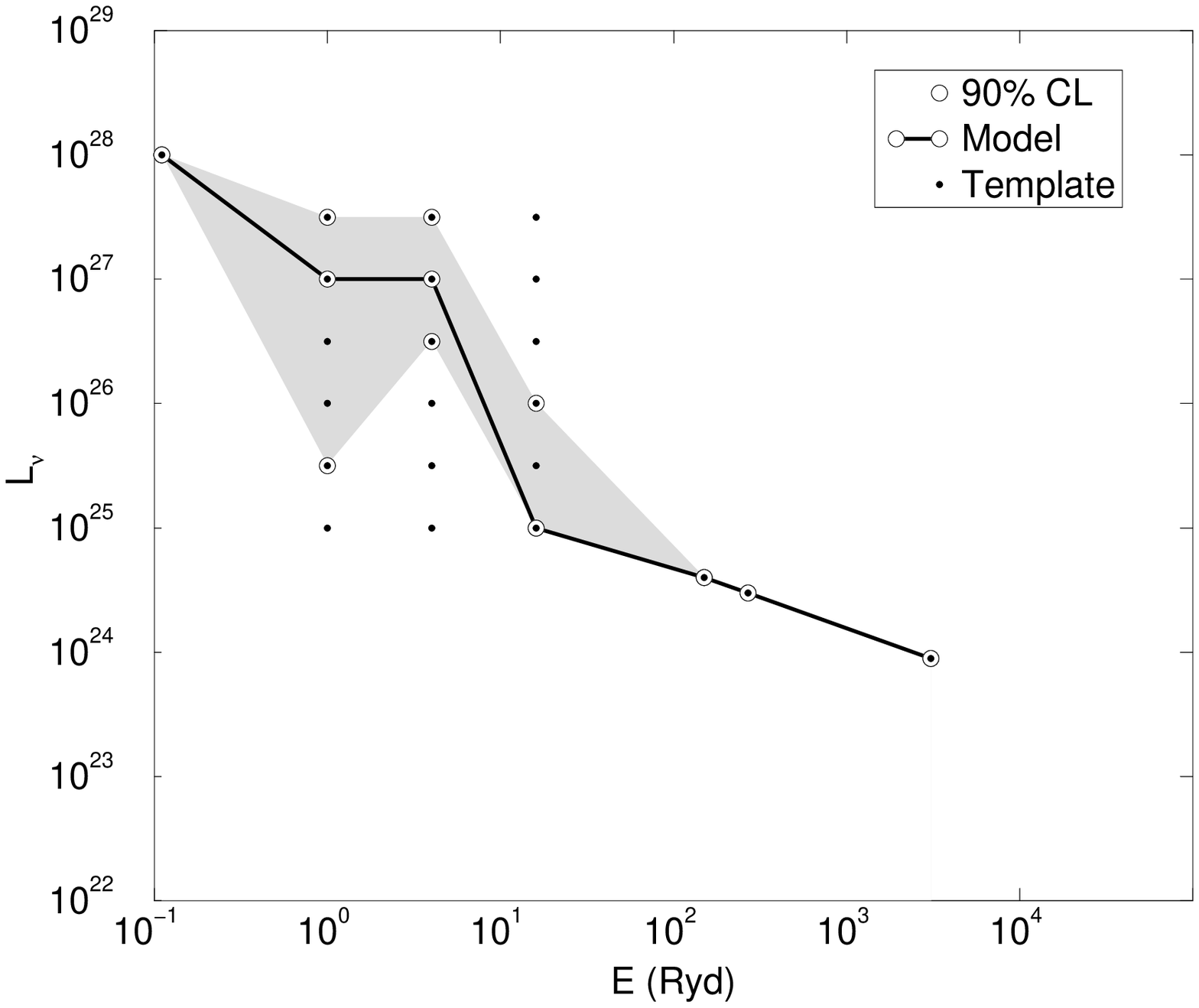} 
\caption{The best-fit SED for the Circinus galaxy with $U=0.45$ and
$F=0.065$ (line). The shaded area is the 90\% confidence limit on the
SED. The black dots are the template SED.
\label{f:best_cir}
}
\end{figure}  

\clearpage
\begin{figure}
\plotone{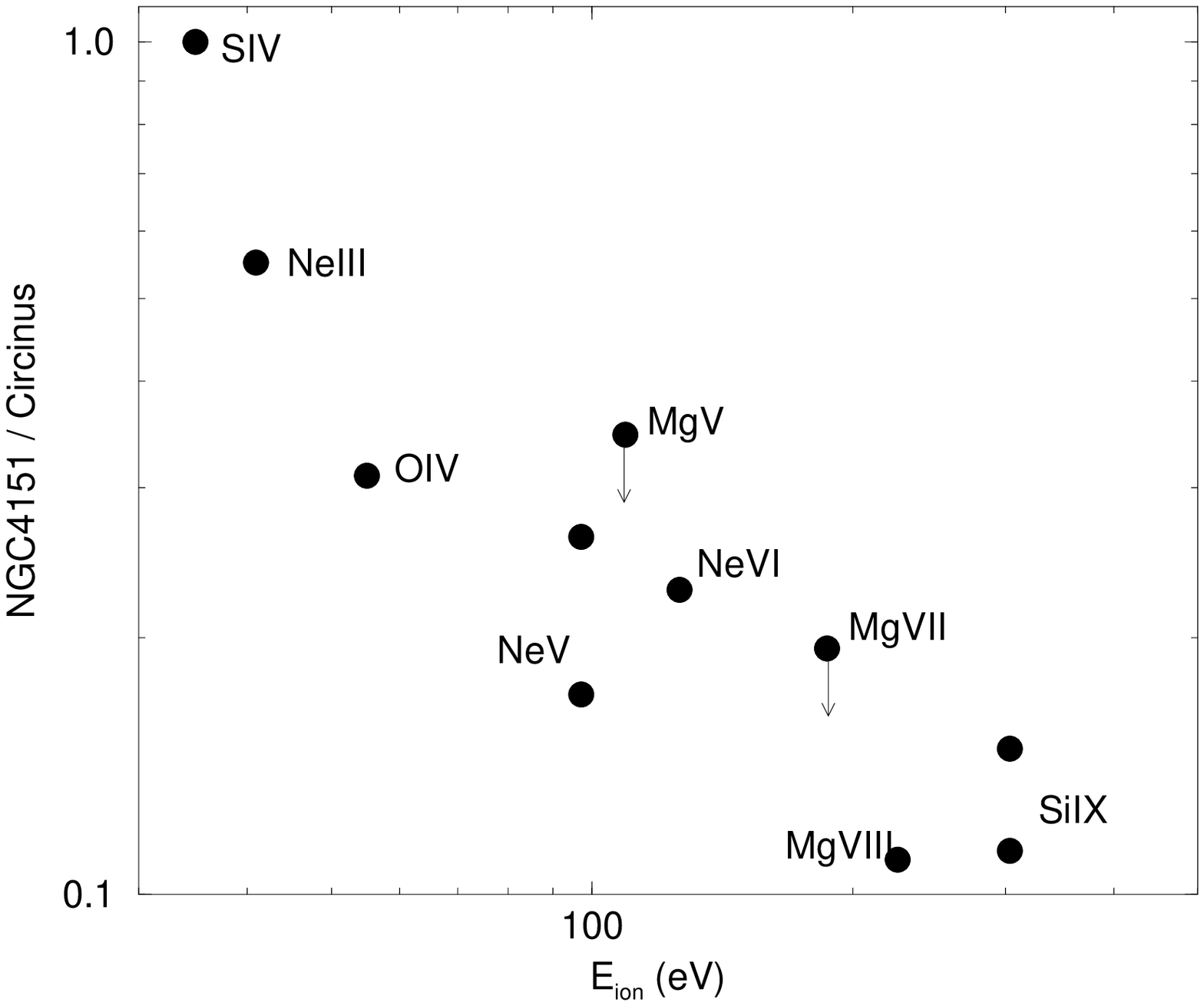} 
\caption{The NGC\,4151 to Circinus ISO lines ratios (normalized to the 
maximal ratio) as function of the ionization energy required to
produce the emitting ion from the preceding ionization stage.
\label{f:cir2ngc}
}
\end{figure} 

\clearpage
\begin{figure}
\plotone{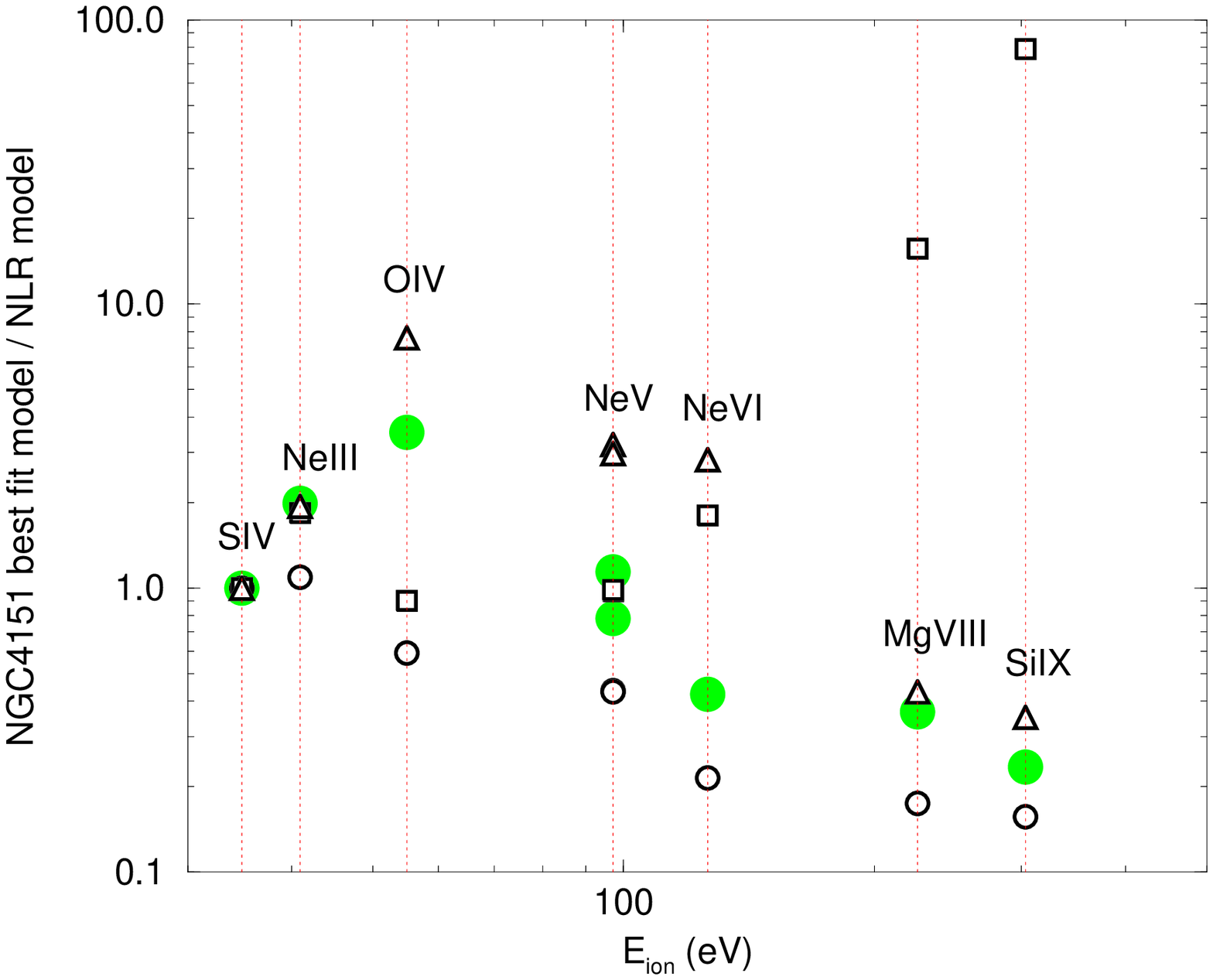}
\caption{The ISO lines ratios between the NGC\,4151 
best full fit model and a sequence of four NLR models. The ratios are
normalized to unity at the $\bSIVb$ line and are shown as function of the
ionization energy required to produce the emitting ion from the preceding
ionization stage. (i) Ratio to the Circinus best fit model (gray
circles). (ii) Ratio to a model identical to the NGC\,4151 best fit model
apart for having the Circinus $U=0.45$ (open circles).  (iii) Ratio to a
model identical to the NGC\,4151 best fit model apart for having the bumpy
SED shown in Figure~\ref{f:abs_SED} (squares). (iv) Ratio to a model
identical to the best full fit NGC\,4151 model, apart for having both
$U=0.45$ and a bumpy SED (triangles). 
\label{f:ngc_models}
}
\end{figure} 

\clearpage
\begin{figure}
\plotone{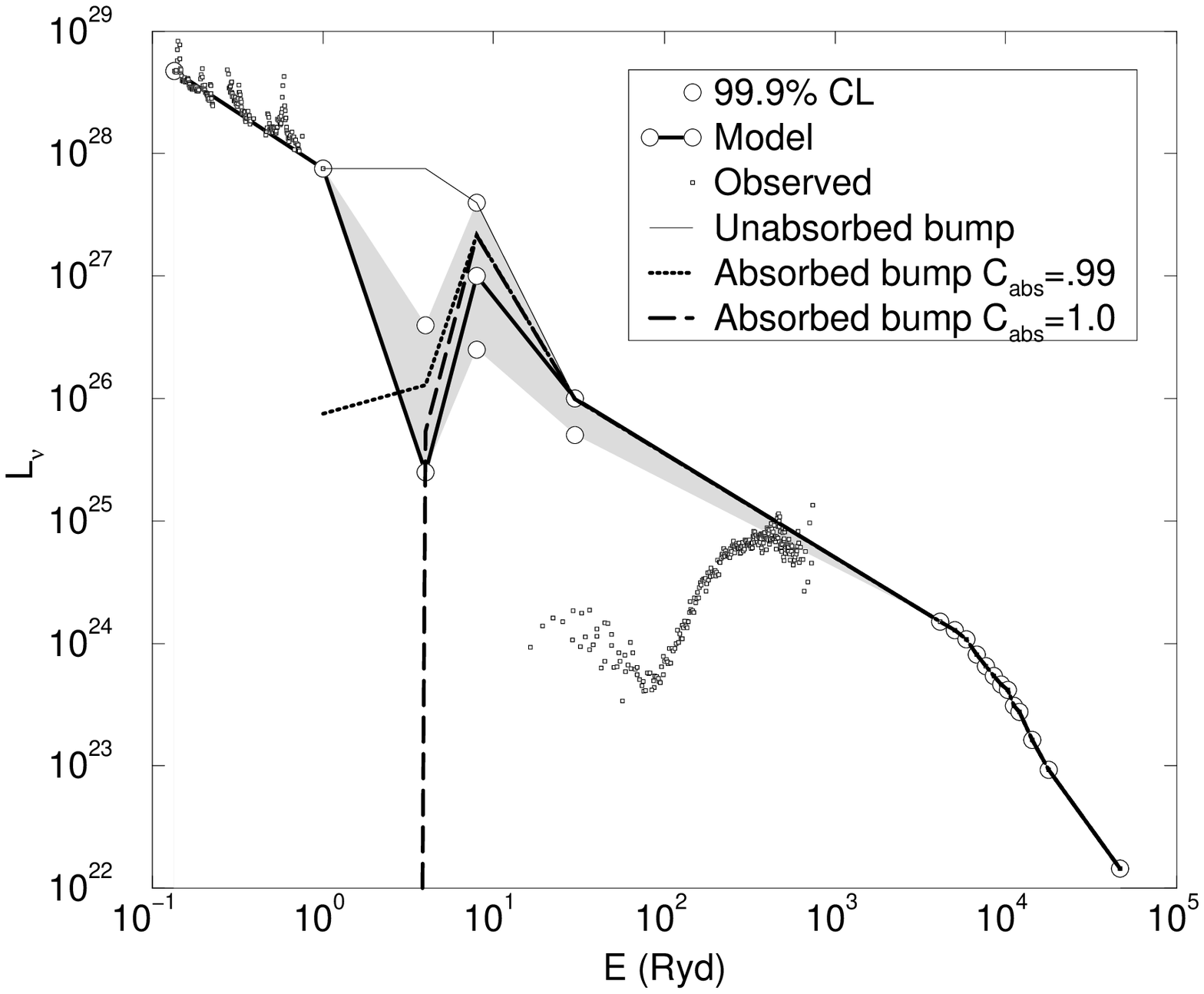}
\caption[absorbed_SED3.eps]{An example of a quasi-thermal bump absorbed
by $5\,10^{19}$\,cm$^{-2}$ of neutral hydrogen. The unabsorbed bump (thin line)
and the absorbed bump with $C_{\rm abs}=0.99$ (dotted line) and $C_{\rm
abs}=1$ (dashed line) are superimposed on the best-fit model based on the
ISO and optical lines ($U=0.025$, $F=6.5\,10^{-4}$, constant
density). 
\label{f:abs_SED}
}
\end{figure} 
\end{document}